\documentclass[twocolumn,showpacs,preprintnumbers,prl]{revtex4}
\usepackage{graphicx,bm,amsmath,amssymb}

\def\gz{\ifmmode{Z\hskip -4.8pt Z}
    \else{\hbox{$Z\hskip -4.8pt Z$}}\fi}

\newcommand{\be}{\begin{equation}}
\newcommand{\ee}{\end{equation}}
\newcommand{\bea}{\begin{eqnarray}}
\newcommand{\eea}{\end{eqnarray}}

\begin{document}

\title{Generalized one-band model based on Zhang-Rice singlets for Tetragonal CuO}
\author{I. J. Hamad}
\affiliation{Instituto de F\'{\i}sica Rosario (CONICET) and Universidad Nacional de Rosario, 
Bv. 27 de Febrero 210 bis, 2000 Rosario, Argentina}
\author{L. O. Manuel}
\affiliation{Instituto de F\'{\i}sica Rosario (CONICET) and Universidad Nacional de Rosario, 
Bv. 27 de Febrero 210 bis, 2000 Rosario, Argentina}
\author{A. A. Aligia}
\affiliation{Centro At\'{o}mico Bariloche and Instituto Balseiro, CNEA, CONICET,  8400 Bariloche, Argentina}
\email{aligia@cab.cnea.gov.ar}

\begin{abstract}
Tetragonal CuO (T-CuO) has attracted attention because of its structure similar to that of the cuprates. 
It has been recently proposed as a compound whose study can give an end to the long debate about the proper microscopic modeling for cuprates. 
In this work, we rigorously derive an effective one-band generalized $t\!-\!J$ model for T-CuO, based on orthogonalized Zhang-Rice singlets, 
and make an estimative calculation of its parameters, based on previous \textit{ab initio} calculations. 
By means of the self-consistent Born approximation, we then evaluate the spectral function and the quasiparticle dispersion for a single hole 
doped in antiferromagnetically ordered half-filled T-CuO. Our predictions show very good agreement with angle-resolved photoemission spectra and 
with theoretical multiband results. We conclude that a generalized $t\!-\!J$ model remains the minimal Hamiltonian for a correct description of 
single-hole dynamics in cuprates.
\end{abstract}

\maketitle

\date{\today }

More than three decades after their discovery, high temperature superconductors still give rise to many debates. 
On the theoretical side, one of the most long-standing and important discussions is about the proper microscopic model for describing 
superconductivity. In this respect and from the outset, attention was focused on the spectral function of a single-hole doped on the 
parent half-filled  compounds, whose quasiparticle (QP) dispersion relation is directly measured in angle-resolved photoemission 
(ARPES) experiments. 
Experimental evidence shows that this doped hole resides on the O 2p$_{\sigma }$ orbitals \cite{nuc,taki,oda}.
For the CuO$_2$ planes that build up the cuprates, Zhang and Rice \cite{ZhangR} proposed that a singlet, called Zhang-Rice (ZR) singlet, 
is formed between the spin of a cooper atom and the spin of the hole residing 
in a linear combination of four ligand oxygen orbitals around the cooper atom. 
Integrating out the oxygen orbitals, a one-band effective model was proposed in which the effective holes 
(representing ZR singlets) reside 
on the cooper atoms and propagate emitting spin excitations, magnons. 
In this model, adding two holes as nearest-neighbors in an antiferromagnetic background costs
less energy than if they are added far apart. This is a simplified view of the pairing glue 
of magnetic origin \cite{fei}. 

Since the proposal of Zhang and Rice, an unclosed debate about the validity of one-band 
effective models has taken place \cite{emery,zhang,ding,bati,Aligia94, Yamase15, Bejas12, Greco09, Chainani17, Adolphs, Ebrahimnejad}. 
Several authors sustain that only the three-band model \cite{eme,varma} is valid 
for describing the physics of the cuprates correctly, where the three bands come from two 
O 2p$_{\sigma }$ orbitals and one Cu 3d$_{x^2-y^2}$ orbital, not only for the insulating parent 
compound at half-filling, but also for many other phases of the rich phase diagram of the cuprates 
and related compounds. 
This issue is of central importance since many investigations have been done in one-band models and hence 
their validity is, at least partially, questioned. 
 
Recently, tetragonal CuO (T-CuO) has been synthesized, by growing epitaxially CuO planes on a substrate 
[(001) SrTiO$_3$] \cite{Siemons-Samal}. T-CuO can be considered as two interpenetrating CuO$_2$ sublattices 
sharing one oxygen atom and hence has two degenerate antiferromagnetic ground states, as shown in Fig. \ref{ordenesmagneticos}. 
ARPES experiments were performed on this compound \cite{Moser}, showing substantial intralayer 
coupling between these two sublattices and a similar dispersion (with some differences) 
to that of the cuprate Sr$_2$CuO$_2$Cl$_2$. 
This material was addressed in a recent work \cite{Adolphs} as a good candidate to discern 
whether one-band models, based on ZR singlets, are valid for describing the physics of CuO planes or if, 
instead, three-band models should be used.      
 
In this Letter, we rigorously derive an effective one-band model for T-CuO and compare its QP 
dispersion with experimental ARPES results and theoretical predictions for the three-band model. 
Using a procedure based on previous derivations of generalized one-band 
effective Hamiltonians \cite{Aligia94}, we start from a spin-fermion model for T-CuO and 
we obtain then its effective one-band model for the ZR singlets. 
The parameters of the model were calculated starting from parameters
determined by constrained-density-functional computations for La$_{2}$CuO$_{4}$ \cite{hyb}, 
and estimating their variations for the T-CuO case. 
We find an effective hopping to first nearest neighbors (NN) between CuO$_2$ sublattices, 
and effective hoppings to first, 
second, and third NN in the same sublattice, 
together with superexchange parameters $J$ (the usual NN antiferromagnetic one for CuO$_2$ planes) 
and a ferromagnetic $J^{\prime}$ (NN in T-CuO, belonging to different CuO$_2$ sublattices). 

Using this model, we calculate the QP dispersion by means of the self-consistent Born approximation (SCBA), 
a reliable and widely used many-body method. 
We compare our results with ARPES experiments in T-CuO, 
obtaining good qualitative and quantitative agreements. 
Our results also recover previous ones from a three-band calculation, 
including particular aspects that were claimed absent in a ZR picture. 
We then conclude that our method is correct for obtaining 
rigorous one-band effective models, and that 
the one-band model that we have derived describes correctly the physics of a single 
doped hole in T-CuO. 

\begin{figure}
{\includegraphics[width=0.35\columnwidth]{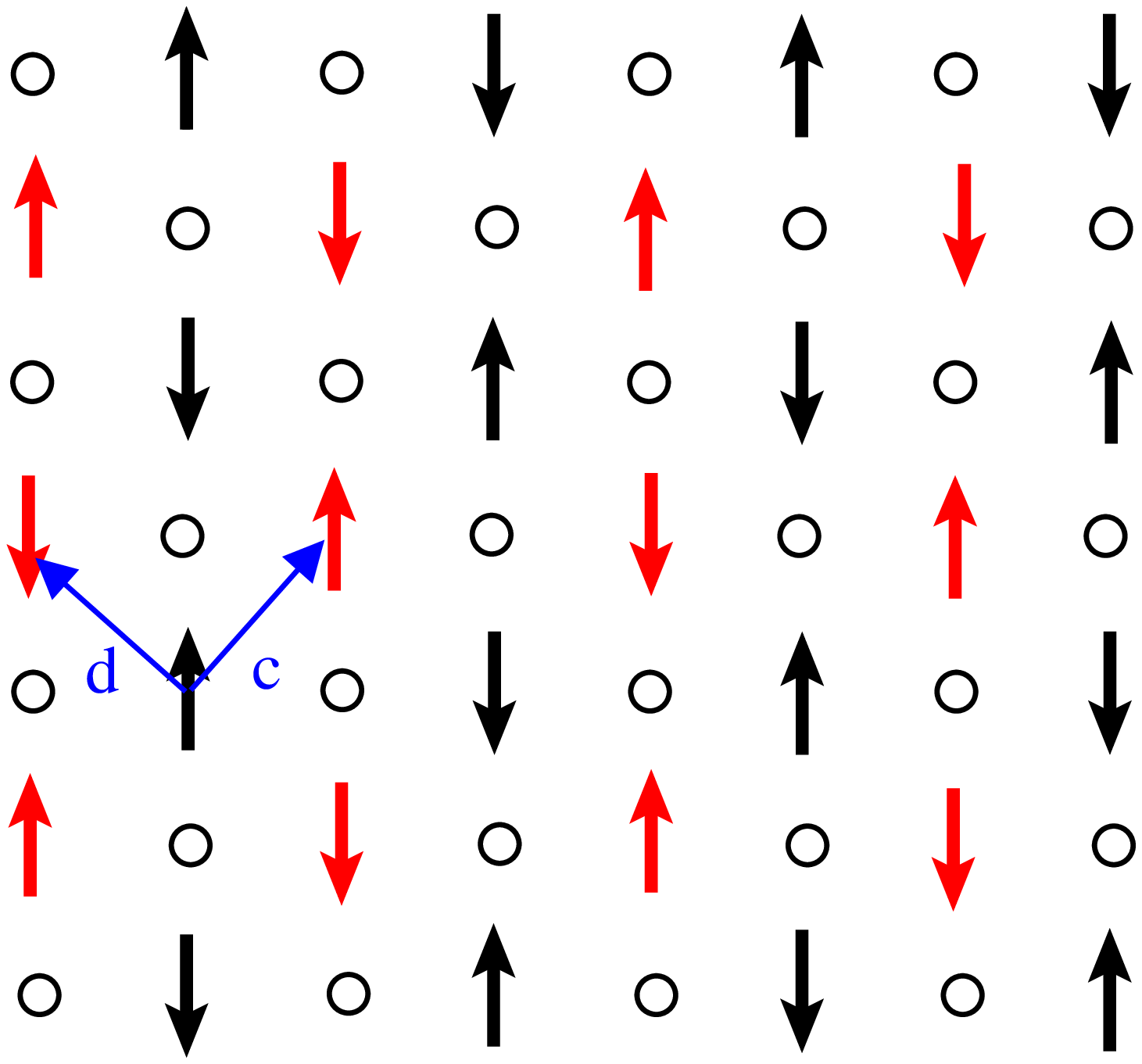}} \hspace{2cm}
{\includegraphics[width=0.35\columnwidth]{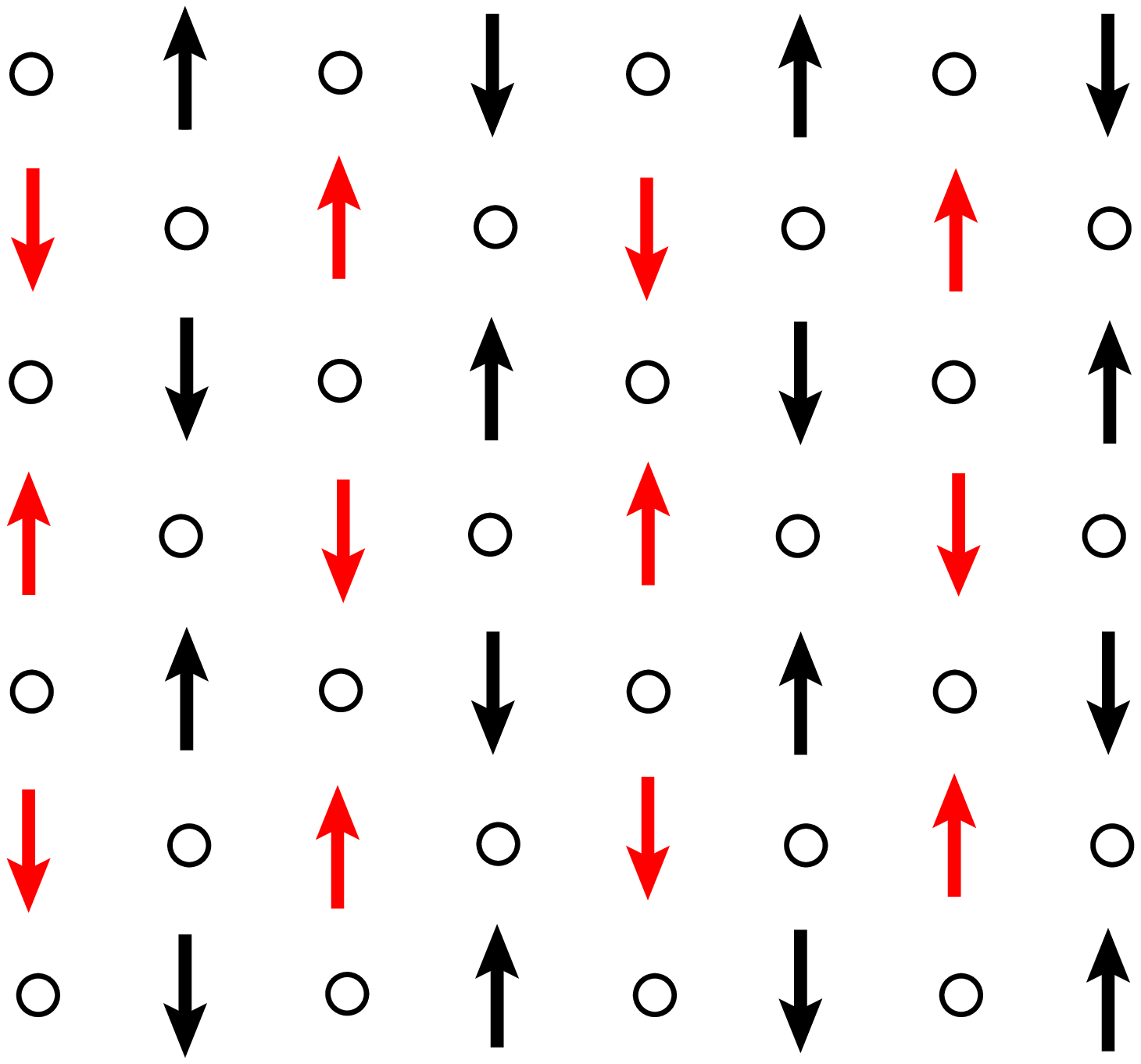}}
\caption{(Color online) The two possible magnetic ground states for T-CuO: 
$\mathbf{Q}=(0,\pi)$ (left) and $\mathbf{Q}=(\pi,0)$ (right). The coordinate versors point in the directions of $\bf{c}$ and $\bf{d}$.  
Arrows indicate spins at Cu sites and circles correspond to the O sites.}
\label{ordenesmagneticos}
\end{figure}

We start from a spin-fermion model (Cu spins and O holes), obtained integrating out 
valence fluctuations at the Cu sites \cite{emery,sf,sf2,Adolphs,supple}. 
With the adequate choice of phases (Fig. S1 of Ref. \onlinecite{supple}) the 
Hamiltonian reads

\begin{eqnarray}
H_{sf}&=&\sum_{i\delta \delta ^{\prime }\sigma }p_{i+\delta ^{\prime
}\sigma }^{\dagger }p_{i+\delta \sigma }\left[ (t_{1}^{sf}+t_{2}^{sf})
(\frac{1}{2}+2\mathbf{S}_{i}\cdot \mathbf{s}_{_{i+\delta }})-t_{2}^{sf}\right] \nonumber \\
&&-J_{d}\sum_{i\delta }\mathbf{S}_{i}\cdot \mathbf{s}_{_{i+\delta
}}
+\frac{J}{2}\sum_{i\delta }\mathbf{S}_{i}\cdot \mathbf{S}_{_{i+2\delta }} \nonumber \\
&&-t_{pp}\sum_{j\gamma \sigma }p_{j+\gamma \sigma }^{\dagger }p_{j\sigma }
+t_{pp}^{\prime }\sum_{j\gamma \sigma }s_{\gamma
}\left( p_{j+\gamma \sigma }^{\dagger }p_{j\sigma }+\mathrm{H.c.}\right) \nonumber \\
&&-\frac{J^{\prime }}{2}\sum_{i\gamma }\mathbf{S}_{i}\cdot \mathbf{S}%
_{_{i+\gamma }},  \label{hsf}
\end{eqnarray}
where $i$ $(j)$ labels the Cu (O) sites and $i+\delta $ ($j+\gamma )$ label
the four O atoms nearest to Cu atom $i$ (O atom $j$). The spin at the Cu
site $i$ (O orbital 2p$_{\sigma }$ at site $i+\delta $) is denoted as $\mathbf{S}_{i}$ 
($\mathbf{s}_{_{i+\delta }}$). The signs $s_{\gamma }=-1$ for
$\gamma \parallel \mathbf{\hat{x}}+\mathbf{\hat{y}%
}$ and $s_{\gamma }=1$ in the perpendicular direction,
being $\mathbf{\hat{x}}$ and $\mathbf{\hat{y}}$ the unit vectors along the directions of 
NN Cu atoms in the CuO$_{2}$ planes (which are second NN in the T-CuO structure).
The parameter  $t_{pp}^{\prime }\simeq 0.6t_{pp}$ (Ref. \onlinecite{Adolphs}). 
This is essentially the same
Hamiltonian as that considered by Adolphs \textit{et al.} \cite{Adolphs}
(we include virtual fluctuations via Cu$^{+3}$) and its low-energy physics
reproduces that of the three-band model \cite{sf2}.
 
Projecting the Hamiltonian over the subspace of orthogonal ZR singlets,  we have derived a one-band 
generalized $t-J$ model 
for T-CuO. All the steps can be found in Ref. \onlinecite{supple}. 
The one-band effective generalized $t-J$ Hamiltonian is: 
\begin{eqnarray}
H_{tJ}^{s}&=&-\sum_{\kappa =0}^{3}t_{\kappa }\sum_{iv_{\kappa }\sigma }\left(
c_{i\sigma }^{\dagger }c_{i+v_{\kappa }\sigma }+\mathrm{H.c.}\right) +\nonumber \\
&&+\frac{J%
}{2}\sum_{iv_{1}}\mathbf{S}_{i}\cdot \mathbf{S}_{_{i+v_{1}}}-\frac{J^{\prime
}}{2}\sum_{iv_{0}}\mathbf{S}_{i}\cdot \mathbf{S}_{_{i+v_{0}}},  \label{hstj}
\end{eqnarray}
where the subscript $\kappa =0$ refers to intersublattice hopping of 
NN Cu atoms in the T-CuO structure, 
while $\kappa =1, 2, 3$, refer to first, second, and
third NN within each CuO$_{2}$ sublattice, respectively. Instead of using arbitrary values 
for the parameters, we have calculated them, keeping the states corresponding to orthogonalized ZR 
singlets and using results from constrained-density-functional calculations \cite{hyb}.
These values are very similar to those corresponding to the model used by Adolphs \textit{et al.} \cite{Adolphs},
as shown in Table 3 of Ref. \onlinecite{supple}. We have checked that the results for both sets are 
quite similar. 
To simplify the discussion we present here only the results 
for the latter. 
The parameters in meV are $t_0=-184$, $t_{1}=369$,  $t_{2}=-11$, $t_{3}=65$, $J=150$, and $J^{\prime}=0$.
This effective model was proposed previously by Moser \textit{et al.} \cite{Moser}.
Here we provide its justification and determine its parameters.

The spectral functions were calculated by means of the SCBA \cite{Martinez91,Lema97,Lema98, Trumper04}, a semianalytic 
method that has been proven to compare very well  
with exact diagonalization (ED) results on finite clusters in different 
systems \cite{Martinez91,Lema97,Trumper04,Hamad08,Hamad12}. 
It is one of the more reliable and checked methods up to date to calculate the hole Green's function, 
and in particular its QP dispersion relation. However, some care is needed to map the 
QP weight between different models \cite{Lema97}. In order to do such calculation, we follow standard 
procedures \cite{Martinez91}. On one hand the magnetic dispersion relation is obtained 
treating the magnetic part of the Hamiltonian at the linear spin-wave level, since the system 
we study has long-range order, and hence its magnetic excitations are semiclassical magnons. 
On the other hand, the electron creation and annihilation operators in the hopping terms are 
mapped into holons of a slave-fermion representation (details in Ref. \onlinecite{supple}). 
Within SCBA, we arrive to an effective Hamiltonian: 
\begin{eqnarray}
H_{\text{eff}} &=&\sum_{\mathbf{k}}\epsilon _{\mathbf{k}}h_{\mathbf{k}}^{\dagger
}h_{\mathbf{k}}+\sum_{\mathbf{k}}\omega _{\mathbf{k}}\theta _{\mathbf{k}%
}^{\dagger }\theta _{\mathbf{k}}+ \nonumber \\
&&+\frac{1}{\sqrt{N}} \sum_{\mathbf{kq}%
}\left(M_{\mathbf{kq}}h_{\mathbf{k}}^{\dagger }h_{\mathbf{k}-\mathbf{q}}\theta _{%
\mathbf{q}}+\mathrm{H.c.}\right), 
\end{eqnarray}

\begin{eqnarray}
\epsilon _{\mathbf{k}} &=&2t_{0}\cos (\mathbf{k}\cdot \mathbf{c})+4t_{2}\cos
(ak_{x})\cos (ak_{y})+\nonumber \\
&&2t_{3}\left[ \cos (2ak_{x})+\cos (2ak_{y})\right] ,  \nonumber \\
\omega_{\bf{k}} &=&\sqrt{A_{\bf{k}}^{2}-4B_{\bf{k}}^{2}}, \nonumber \\
M_{\mathbf{kq}} &=&2t_{0}\left\{ \cos \left[ (\mathbf{k-q})\cdot \mathbf{c}%
\right] u_{\mathbf{q}}-\cos (\mathbf{k}\cdot \mathbf{c})v_{\mathbf{q}%
}\right\} +\nonumber \\
&&2t_{1}\left[ u_{\mathbf{q}}\zeta (\mathbf{k-q})-v_{\mathbf{q}%
}\zeta (\mathbf{k})\right],   
\label{HeffSCBA}
\end{eqnarray}

\normalsize

\noindent where $\epsilon _{\mathbf{k}}$ is the bare hole dispersion (with no coupling to magnons), 
$\omega_{\bf{k}}$ is the magnon dispersion relation, with 
$A_{\bf{k}} =2J-J^{\prime }\cos (\bf{c}\cdot \bf{k})$, 
$B_{\bf{k}}=\frac{J}{4}\sum_{v_{1}}\cos (v_{1}\cdot \bf{k})-\frac{J^{\prime }}{2}\cos (\bf{d}\cdot \bf{k})$, 
and $M_{\mathbf{kq}}$ is the vertex that couples the hole with magnons. 
Here $\zeta (\mathbf{k}) =\cos (ak_{x})+\cos (ak_{y})$, and 
$\mathbf{c}=b(\mathbf{\hat{x}}+\mathbf{\hat{y}})$, $\mathbf{d}=b(\mathbf{-\hat{x}}+\mathbf{\hat{y}})$, 
being $a=2b$ the distance between Cu atoms in the CuO$_{2}$ planes. 
The vectors $\bf{c}$ and $\bf{d}$ are indicated in Fig. \ref{ordenesmagneticos}.
\begin{figure}[h]
\vspace{1cm}
\includegraphics[scale=0.35]{reldisp_Moser_2ordenes_2.eps}%
\begin{picture}(0,0)
\put(-205,90){\includegraphics[width=3.7cm]{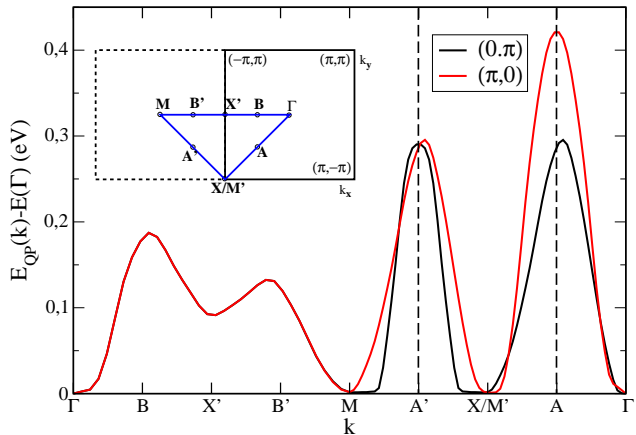}}
\end{picture}
\caption{(Color online) Quasiparticle dispersion relation (relative to $\Gamma$) along the path marked in the inset, 
the same as the one measured in the ARPES experiment in Ref. \onlinecite{Moser}. 
A broadening equivalent to 20 meV was applied to the spectral functions (see text).}
\label{reldispMoser}
\end{figure}
We now compare our results with ARPES experiments performed on T-CuO, specifically with the those in Figs. 2 and 3 of Ref. \onlinecite{Moser}. 
For that purpose, we adopt in Figs. \ref{reldispMoser} and  \ref{setB}, an electron picture. 
In Fig. \ref{reldispMoser} we show the QP dispersion derived from our SCBA calculation. 
This should be compared with the blue points in Fig. 2 of Ref. \onlinecite{Moser}, and also with the white 
points in the same figure, corresponding to exact diagonalization of a one-band Hubbard model in 20 sites. 
In our calculation, a broadening equivalent to 20 meV (controlled by means of the parameter 
$\delta$ in Eq. \ref{HeffSCBA}), similar to the experimental resolution (30 meV \cite{Moser}), 
was applied to the spectral functions. Taking into account the two possible magnetic ground states for T-CuO, we obtain the 
two QP dispersions shown in Fig. \ref{reldispMoser}. It can be observed that  the dispersion corresponding to ${\bf{Q}}=(\pi,0)$ 
recovers all the main features of the experimental dispersion, and hence our results can distinguish between the possible degenerate 
magnetic orders in the experiment. In particular, we recover the asymmetry between the  points $\Gamma$ and $X'$, $B$ and $B'$, 
and $A$ and $A'$. Moreover, we obtain, $E(A)-E(A')=128$ meV, $E(B)-E(B')=64$ meV, and $E(\Gamma)-E(M)= 10$ meV, while the experimentally 
measured energy differences are $140$ meV, $60$ meV, and $180$ meV, respectively \cite{Moser}. 
The agreement is very good, except in the last case. 
This discrepancy is quite likely due to missing quasiparticle peaks with small weight in the experiment (see also Fig. S4 of Ref. \cite{supple}).
In that sense, we note that the $\Gamma$ point (and points located in its vicinity) shows a very broad spectrum (see Figs. 2 and 3 in Ref. \cite{Moser} ), 
and hence there may be some uncertainty in the determination of the QP energy which could explain this discrepancy. 
The bandwidth of the 
QP dispersion, along this path, taken from our SCBA calculation is $0.3 eV$, very similar to the bandwidth of the experimental 
dispersion, approximately $0.4 eV$.   
\begin{figure}[h]
\vspace{1cm}
\includegraphics[width=8.cm]{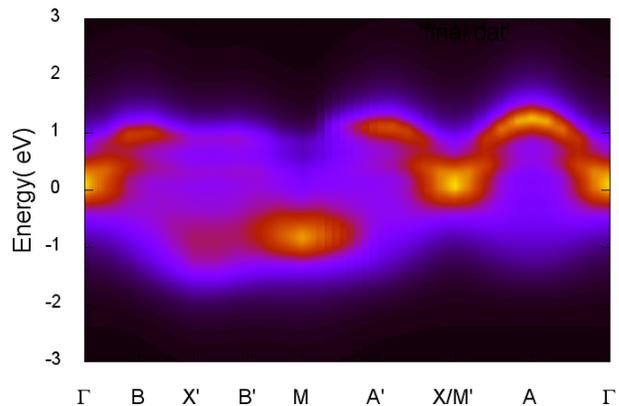}
\caption{(Color online) SCBA intensity map along the same path as in Fig. \ref{reldispMoser}. 
The assumed magnetic order is $(\pi,0)$.}
\label{setB}
\end{figure}
\begin{figure*}
{\includegraphics[width=1.5\columnwidth]{reldisp_Adolphs_paper.eps}} \hspace{0.3cm} \centering 
\raisebox{0.12\height} {\includegraphics[width=0.45\columnwidth]{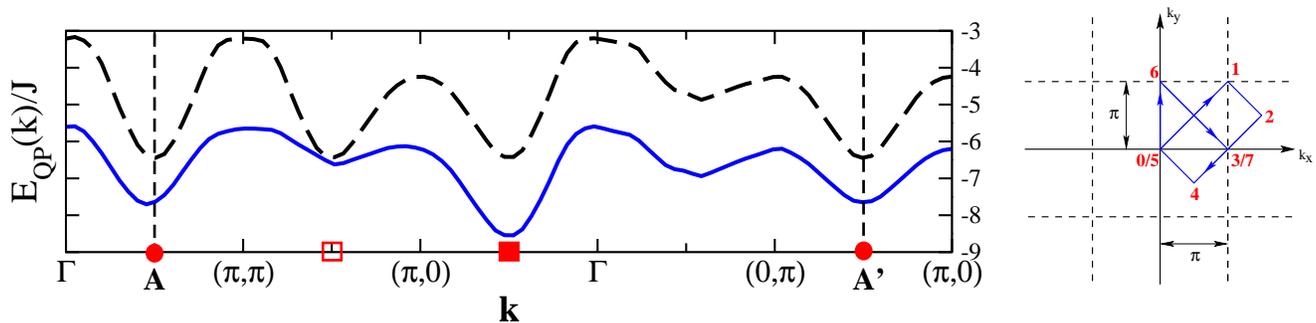}}
\caption{(Color online) SCBA hole's dispersion relation in units of $J$ (0.15 eV) along the path marked in the inset.
Black dashed line: result corresponding to $t_0=0$ (decoupled sublattices). 
Blue full line: full result with $t_0=-184$ meV.}
\label{reldispAdolphs}
\end{figure*} 
We have also calculated an intensity curve along the same path as in the experiment, 
to compare with the ARPES intensities (Fig. 2 of Ref. \onlinecite{Moser}). We show only the intensity corresponding to ${\bf{Q}}=(\pi,0)$, 
since for this order our QP dispersion recovers the experimental one. 
For this calculation a broadening equivalent to 170 meV was applied, in order to make the intensity plot softer. 
The results are shown in Fig. \ref{setB}. The similarities with the experimental curve 
follow the trends explained in the previous paragraph. It is worth to mention that, on one hand, 
at some points in the experimental curve the effect of the 
ARPES matrix elements is very strong, especially around the $X/M'$ point, where there is no 
intensity at all in the ARPES data, and on the other hand a $\beta$ band seems to merge with the 
QP band, specially at the $X/M'$ point but also possibly around the $M$ point. 
So at these two points, in particular around the $X/M'$ points, the comparison of our calculation 
with the experiment is obscured by these experimental facts. 
Finally, it is worth to mention that in the case that the illuminated area in the 
ARPES experiments contains domains with both magnetic  
${\bf{Q}}=(\pi,0)$ and ${\bf{Q}}=(0,\pi)$ vectors (as mentioned above, they are degenerate), 
the QP dispersion should be a superposition of both curves shown in Fig. \ref{reldispMoser}, 
which does not seem to be what is observed in the experiment \cite{aclaracion}. 
The intensity curve Fig. \ref{setB} should also change accordingly, but in our case we have checked that 
the only noticeable changes occur around the $X/M'$ point, at which nevertheless there is no intensity in the 
ARPES data corresponding to the band ascribed to ZR singlets \cite{Moser}.

In general, the spectral function corresponding to a definite momentum contains, in the hole picture, a low energy pole, 
whose energy defines the QP energy, and a high energy part which is related to the incoherent movement of the hole, having its origin in multimagnon processes \cite{Hamad08}. 
When the quasiparticle weight is significant, the brighter areas in Fig. \ref{setB} will coincide with the energy of the QP in Fig. \ref{reldispMoser}. 
On the contrary when the incoherent part of the spectral function takes most of the spectral weight, this will not happen. Points like $\Gamma$ and $M$ have low QP weight, 
while on the contrary for the lines $B-B'$, $A'-A$ the QP weight is relatively high (some spectral functions can be seen in Fig S4 of ref. \onlinecite{supple}).

It was claimed previously that the one-hole dispersion in T-CuO requires a three-band model to be described 
correctly \cite{Adolphs}. 
The evidence presented came from a variational calculation on the spin-fermion model Eq. (\ref{hsf}), 
whose results a one-band model 
supposedly cannot capture. In particular, it was shown that the minimum that the QP dispersion has at 
$(\pi/2,\pi/2)$ for CuO$_2$ (or, equivalently in T-CuO, if the two CuO$_2$ sublattices are disconnected), shifts 
along the diagonal $\Gamma \equiv (0,0)-(\pi,\pi)$, towards the $\Gamma$ point, 
when the two CuO$_2$ sublattices are connected to form T-CuO. This happens for ${\bf{Q}}=(0,\pi)$.  Alternatively, the shift is along the antidiagonal towards $X/M$ for ${\bf{Q}}=(\pi,0)$. 
This is what we have shown in figure \ref{reldispMoser}.    
These results are in line with previous investigations for CuO$_2$ planes \cite{Ebrahimnejad}, 
where it was claimed that a one-band $t-t'-t''-J$ model has a minimum at $(\pi/2,\pi/2)$ that along the 
diagonal of the Brillouin zone is controlled by spin fluctuations, while in the three-band model the variational 
method used in Ref. \onlinecite{Ebrahimnejad} 
does not need to include spin fluctuations in order to have an 
absolute minimum at $(\pi/2,\pi/2)$. 
 
Using the generalized $t-J$ model [Eq. (\ref{hstj})] derived from $H_{sf}$ [Eq. (\ref{hsf})] 
we now calculate the QP dispersion along the same path as in Ref. \onlinecite{Adolphs}
and with the corresponding parameters (set B of Table III of Ref. \onlinecite{supple}), and ${\bf{Q}}=(0,\pi)$.  
Results are shown in Fig. \ref{reldispAdolphs}, plotted adopting the hole's picture. 
As before, a broadening equivalent to 20 meV was applied to the spectral functions, 
but the results do not depend significantly on this (unless broadenings an order of magnitude larger are applied). 
It is clear that when both sublattices are connected through the $t_0$ term, the QP dispersion relation derived from 
$H_{sf}$ is recovered. In particular, we obtain a shift of the QP minimum along the diagonal towards 
the $\Gamma$ point, although this shift is lower (about half) in magnitude than the one obtained with the three-band model. 
This difference might be due to the different theoretical treatments used by Adolphs \textit{et al.}
to solve $H_{sf}$ [Eq. (\ref{hsf})] and by us to solve $H_{tJ}$ [Eq. (\ref{hstj})] 
In this respect, we remark it is very difficult to decide which theoretical treatment gives more accurate results 
from quantitative differences of 
this kind, since on one hand both compare very well with ED results in finite clusters, while on the other hand no 
experiment so far could even measure this shift in the QP dispersion relation. 
We also remark that varying $t_2$, the QP dispersion relation is not changed apart from a constant shift 
(in agreement with previous results \cite{Ebrahimnejad}). This is important since $t_2$ is the parameter 
obtained with less accuracy.

The shift in our model is not caused by the coupling of the hole with spin fluctuations, which in fact conspires against it. 
This can be seen from the effective Hamiltonian Eq. (\ref{HeffSCBA}), since the bare-hole dispersion 
({\it i.e.} with no coupling to magnons) 
$\epsilon _{\mathbf{k}} = 2t_{0}\cos (\mathbf{k}\cdot \mathbf{c})
+4t_{2}\cos(ak_{x})\cos (ak_{y})+ 2t_{3}\left[ \cos (2ak_{x})+\cos (2ak_{y})\right]$ has a minimum, 
along the diagonal $k_x=k_y$, that shifts  from $(\pi/2,\pi/2)$ towards the $\Gamma$ point when the 
intersublattice hopping $t_0$ is turned on. For example, the bare hole minimum is at $(0.4 \pi,0.4\pi)$ for the 
parameter set we used. However, when the interaction of the bare hole with spin fluctuations (magnons) is taken into account 
through the vertex $M_{kq}$, the minimum shifts back towards $(\pi/2,\pi/2)$. The shift obtained is about $10\%$ of the distance between the $A$ and $\Gamma$ points.  
Note that the SCBA contains an infinite number of spin fluctuations while only a few are included in 
the treatment of Ref. \onlinecite{Adolphs}.
In any case, we have shown that a ZR one-band model 
can explain a shift in the QP minimum at $(\pi/2,\pi/2)$, and that the interaction of the bare hole with spin fluctuations is 
not responsible for this shift. Finally, the QP bandwidth along this path is, in our one-band model, of the order 
of $3J$, slightly less than the result 
from the variational method in the three-band model Eq. (\ref{hsf}) \cite{Adolphs}.   

Overall, we conclude that our effective generalized one-band model, rigorously derived from orthogonalized 
Zhang-Rice singlets, and without free parameters, not only does recover characteristics of the three-band model, 
but also its predictions agree qualitatively and quantitatively with ARPES experiments in tetragonal CuO. 
 
{\acknowledgments
 We thank A. Greco for fruitful discussions. IJH was partially supported by PICT-2014-3290. IJH and LOM are partially supported by PIP 0364 of CONICET.
AAA is sponsored by PIP 112-201101-00832 of CONICET and PICT 2013-1045 of the ANPCyT.}

\pagebreak
%\widetext
\onecolumngrid

\begin{center}
\textbf{\large{Supplemental Material: Generalized one-band model based on Zhang-Rice singlets for Tetragonal CuO}}

\vspace{0.8cm}

I. J. Hamad, L. O. Manuel, A. A. Aligia
\end{center}

\setcounter{equation}{0}
\setcounter{figure}{0}
\setcounter{table}{0}
\setcounter{page}{1}
\makeatletter

\begin{widetext}

\section{The starting model}

\label{sf}

\begin{figure}[h]
\includegraphics[width=8. cm]{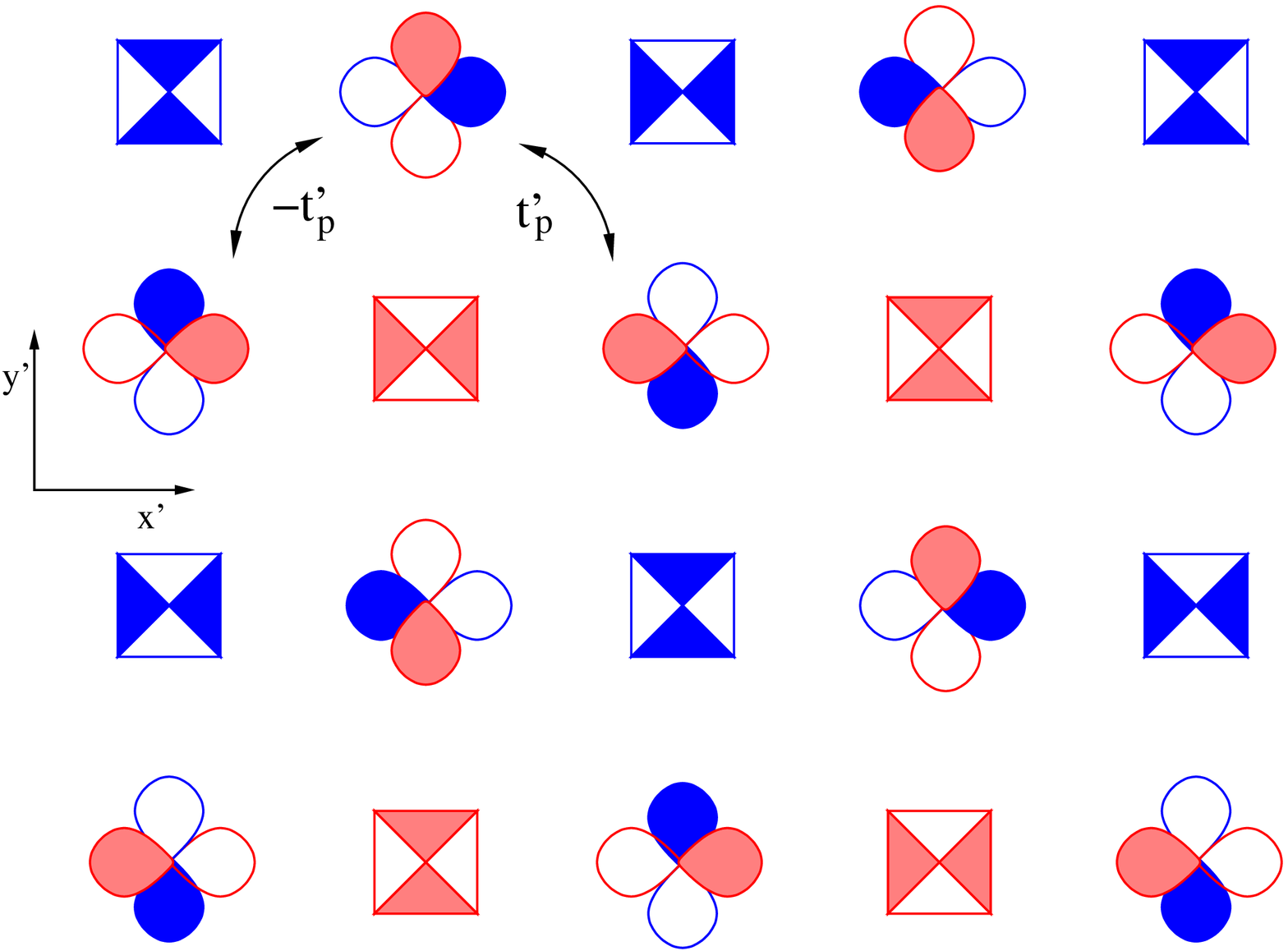}
\caption{Scheme of the 3d$_{x^{2}-y^{2}}$ (squares) and 2p$_{\sigma }$ (number 8) orbitals of the T-CuO planes.
Blue and red orbitals belong to different CuO$_2$ sublattices. Empty and filled parts of the
orbitals have opposite signs.}
\label{fases}
\end{figure}

It is known that for energies below 1 eV, the physics of the superconducting
cuprates is described by the three-band Hubbard model $H_{3b}$, which
contains the 3d$_{x^{2}-y^{2}}$ orbitals of Cu and the 2p$_{\sigma }$
orbitals of O \cite{varma,eme,note}. We denote by 
$\mathbf{\hat{x}}$ and $\mathbf{\hat{y}}$ the unit vectors along the directions of nearest-neighbor
(NN) Cu atoms in the CuO$_{2}$ planes (which are second NN in the T-CuO
structure) and $a$ their distance. Experimental evidence about the symmetry
of holes in cuprate superconductors \cite{nuc,taki,oda} shows that the
undoped system has one hole in each Cu 3d$_{x^{2}-y^{2}}$ orbital, so that
all Cu atoms are in the oxidation state 2+, while added holes enter the O 2p$%
_{\sigma }$ orbitals. Therefore, it is natural to eliminate the Cu-O hopping 
$t_{pd}$ and the states with Cu$^{+}$ and Cu$^{+3}$ (keeping them as virtual
states) by means of a canonical transformation \cite{eme2,sf}. The resulting
effective Hamiltonian, which consists of Cu 1/2 spins and O holes is usually
called spin-fermion model. As usual, we change the phases of half the Cu and
O orbitals so that the Cu-O hopping has the same sign independent of
direction (see Fig. \ref{fases})

\begin{equation}
d_{i\sigma }\longrightarrow e^{i\mathbf{Q}\cdot (\mathbf{R}_{i}-\mathbf{R}%
_{i}^{0})}d_{i\sigma },\text{ }p_{j\sigma }\longrightarrow e^{i\mathbf{Q}%
\cdot (\mathbf{R}_{j}-\mathbf{R}_{j}^{0})}p_{j\sigma },\text{ }\mathbf{Q=}%
\frac{\pi }{a}\left( \mathbf{\hat{x}}-\mathbf{\hat{y}}\right),  \label{phases}
\end{equation}%
where $\mathbf{R}_{i}^{0}$ ($\mathbf{R}_{j}^{0}$) is a fixed Cu (O)
position. After this transformation, for one hole added to the undoped
system, the model can be written as \cite{sf,sf2}

\begin{equation}
H_{sf}^{p}=\sum_{i\delta \delta ^{\prime }\sigma }p_{i+\delta ^{\prime
}\sigma }^{\dagger }p_{i+\delta \sigma }\left[ (t_{1}^{sf}+t_{2}^{sf})(\frac{%
1}{2}+2\mathbf{S}_{i}\cdot \mathbf{s}_{_{i+\delta }})-t_{2}^{sf}\right]
-J_{d}\sum_{i\delta }\mathbf{S}_{i}\cdot \mathbf{s}_{_{i+\delta
}}-t_{pp}\sum_{j\gamma \sigma }p_{j+\gamma \sigma }^{\dagger }p_{j\sigma }+%
\frac{J}{2}\sum_{i\delta }\mathbf{S}_{i}\cdot \mathbf{S}_{_{i+2\delta }}.
\label{hsf}
\end{equation}%
Here $i$ $(j)$ labels the Cu (O) sites and $i+\delta $ ($j+\gamma )$ label
the four O atoms nearest to Cu atom $i$ (O atom $j$). The spin at the Cu
site $i$ (O orbital 2p$_{\sigma }$ at site $i+\delta $) is denoted as $%
\mathbf{S}_{i}$ ($\mathbf{s}_{_{i+\delta }}$). The first term corresponds to
an effective O-O hopping with possible spin flip with a Cu spin, $t_{1}^{sf}$
($t_{2}^{sf}$) correspond to virtual processes through Cu$^{+}$ (Cu$^{+3}$).
When both NN vectors coincide ($\delta =\delta ^{\prime }$), the second term
contains a term of the form of the second one. The total Cu-O NN exchange is 
$J_{K}=2(t_{1}^{sf}+t_{2}^{sf})-J_{d}$. In second-order perturbation theory, 
$J_{d}$ vanishes if the on-site O repulsion is neglected \cite{eme2,sf}, but
in general $J_{d}>0$. The third term is the direct O-O hopping and the last
one is the exchange between nearest Cu atoms.

It has been shown that $H_{sf}^{p}$ with parameters slightly renormalized to
fit the energy levels of a CuO$_{4}$ cluster in some symmetry sectors
(solving small matrices) reproduces Cu and O photoemission and inverse
photoemission spectra and spin-spin correlations functions of the three band
model $H_{3b}$ in a Cu$_{4}$O$_{8}$ cluster \cite{sf2}. This was later
extended to angle-resolved Cu and O photoemission intensities \cite{erol}.
Therefore we assume that $H_{sf}^{p}$ is an accurate representation of the
low-energy physics of $H_{3b}$.

The system of tetragonal CuO (T-CuO) consists of two interpenetrating CuO$%
_{2}$ sublattices, one displaced with respect to the other in a vector $%
\gamma $ connecting two NN O ions (see Fig. \ref{fases}). One of the sublattices can
be described by Eq. (\ref{hsf}). The O orbitals of the other sublattice lie
on the same site as the previous ones but are orthogonal to them. We
label $i^{\prime }$ the Cu orbitals of the second sublattice and $%
q_{i^{\prime }+\delta \sigma }$ the annihilation operators of the four O 2p$%
_{\sigma }$ orbitals nearest to Cu site $i^{\prime }$. The Hamiltonian that
describes the second sublattice $H_{sf}^{q}$, has the same form as $%
H_{sf}^{p}$ with $i$ replaced by $i^{\prime }$ and the O $p$ operators by
the $q$ ones. Including the NN O-O hopping and the NN Cu-Cu exchange between
both sublattices, the Hamiltonian reads

\begin{equation}
H_{sf}=H_{sf}^{p}+H_{sf}^{q}+t_{pp}^{\prime }\sum_{j\gamma \sigma }s_{\gamma
}\left( p_{j+\gamma \sigma }^{\dagger }q_{j\sigma }+\mathrm{H.c.}\right) -%
\frac{J^{\prime }}{2}\sum_{i\gamma }\mathbf{S}_{i}\cdot \mathbf{S}%
_{_{i+\gamma }},  \label{h}
\end{equation}%
where $s_{\gamma }=-1$ for $\gamma \parallel \mathbf{\hat{x}}+\mathbf{\hat{y}%
}$ and $s_{\gamma }=1$ in the perpendicular direction (see Fig. \ref{fases}) and 
$t_{pp}^{\prime }\simeq 0.6t_{pp}$ \cite{adol}. This is essentially the same
Hamiltonian as that considered by Adolphs \textit{et al} \cite{adol}. The
last term is originated by perturbation theory in fourth order in the Cu-O
hopping $t_{pd}$ involving two O atoms, each one forming a Cu-O-Cu angle of
90 degrees, and virtual states with an O occupied by two holes in
perpendicular orbitals (one $p_{j\sigma }$ and one $q_{j\sigma ^{\prime }}$%
). It is ferromagnetic due to the Hund rules at the O atoms. Estimating the
difference between singlet and triplet two-hole states from that between $%
^{1}$D and $^{3}$P states in atomic O (1.97 eV \cite{moore}) and taking the
rest of the parameters from constrained-density-functional calculations for
La$_{2}$CuO$_{4}$ (Ref. \onlinecite{hyb}) we obtain $J^{\prime }=2.7$ meV.
This value is very sensitive to the Cu-O charge transfer energy $\Delta $\
and to the Coulomb repulsion $U_{pd}$ between Cu and O. For example changing 
$U_{pd}$ from 1.2 eV to 0, $J^{\prime }$ increases to 34 meV.

In Table \ref{parsf} we show an estimation of the parameters of $H_{sf}$
based on previous results \cite{erol} of the low-energy reduction procedure
from $H_{3b}$ with parameters derived from constrained-density-functional
calculations (set A) and the parameters used by Adolphs \textit{et al}. \cite{adol} (set B).

Since the structure of T-CuO is different from that of the cuprates, the
estimation of the parameters is very approximate. It would be desirable to
have estimations for the parameters of $H_{3b}$ for T-CuO, in particular the
charge-transfer energy $\Delta$. In absence of them one can estimate the
hopping terms taking into account that the CuO distance is increased from
the value $b=a/2=1.895$ \AA\ used in Ref. \onlinecite{hyb} to $b=1.9525$ 
\AA\ in T-CuO \cite{supm}, using the scaling $t_{pd} \propto d^{-7/2}$, 
$t_{pp} \propto d^{-2}$ for the dependence on the distance $d$ of the hopping
parameters \cite{harr}. 
This leads to a reduction of $t_{pp}$ by a factor 0.94 and using that for small 
$t_{pd}$,  $t_{i}^{sf} \propto t^2_{pd}$ a reduction of these effective hoppings by a factor 
0.81 might be expected, neglecting the influence of the change in on-site energies and repulsions.

\begin{table}[h]
\caption{Parameters of the spin-fermion model for T-CuO in eV.}
\label{parsf}%
\begin{ruledtabular}
\begin{tabular}{llllllll}
set & $t_{1}^{sf}$ & $t_{2}^{sf}$ & $J_d$ & $t_{pp}$ & $t_{pp}^{\prime }$ & $J$ & $J^{\prime}$ \\ 
\hline
A  & 0.37 & 0.08 & 0.28 &  0.56  & 0.336   & 0.13 & 0.0027 \\
B  & 0.45 & 0    & 0.48 &  0.615 & 0.369 & 0.15 & 0       \\

\end{tabular}
\end{ruledtabular}
\end{table}

\section{The generalized $t-J$ model for CuO$_{2}$ planes.}
\label{gtj}

Zhang and Rice proposed that the low-energy physics of the cuprates is
dominated by the now called Zhang-Rice singlets (ZRS) \cite{zr}. In the
language of $H_{sf}^{p}$, for which fluctuations via Cu$^{+}$ and Cu$^{+3}$
are included virtually, for each Cu site $i$ these singlets have the form

\begin{eqnarray}
|i\tilde{s}\rangle &=&\frac{1}{\sqrt{2}}\left( \tilde{\pi}_{i\uparrow
}^{\dagger }d_{i\downarrow }^{\dagger }-\tilde{\pi}_{i\downarrow }^{\dagger
}d_{i\uparrow }^{\dagger }\right) |0\rangle ,\text{ }  \label{zr} \\
\text{ }\tilde{\pi}_{i\sigma } &=&\frac{1}{2}\sum_{\delta }p_{i+\delta
\sigma },  \label{pino}
\end{eqnarray}%
where $d_{i\sigma }^{\dagger }$ creates a hole at the 3d$_{x^{2}-y^{2}}$
orbital of site $i$. Retaining only ZRS and neglecting the rest of the
states (or including them perturbatively) and mapping these states $|i\tilde{%
s}\rangle \leftrightarrow |i0\rangle $ to the vacuum at site $i$ (which
corresponds to a full 3d shell) leads to a one-band generalized $t-J$ model.
Several systematic studies of this mapping were made starting for either $%
H_{3b}$ or $H_{sf}^{p} $,\ which include more terms than just the NN hopping 
$t$ and the exchange $J $. See for example Refs. \cite{sys,bel,fei}. A
difficulty with the states $|i\tilde{s}\rangle$ is that they have a finite
overlap for NN Cu sites $i$ and $i+2\delta $. Using these non-orthogonal
singlets Zhang proved that the mapping from $H_{sf}^{p}$ to the $t-J$ model
is exact for $t_{1}^{sf}=t_{pp}=0.$\cite{zhang} This procedure was
generalized to include the other terms of $H_{sf}^{p}$ , leading to
additional terms in the generalized $t-J$ model \cite{sys}.

However, orthogonalizing the states leads to a simpler mapping procedure
which is in general preferred and is more accurate when 
$t_{1}^{sf}>t_{2}^{sf} $ (fluctuations via Cu$^{+}$ dominate) \cite{sys},
which is in general the case. The trick to obtain orthonormal states is to
transform Fourier the $\tilde{\pi}_{i\sigma }$ operators, normalize in
wave-vector space, and transform back \cite{zr}, leading to

\begin{equation}
\pi _{i\sigma }=\frac{1}{N}\sum_{\mathbf{k}}e^{-i\mathbf{k\cdot R}_{i}}\beta
_{\mathbf{k}}\sum_{m}e^{i\mathbf{k\cdot R}_{m}}\tilde{\pi}_{m\sigma },\text{ 
}\beta _{\mathbf{k}}=\left[ \cos ^{2}(k_{x}b)+\cos ^{2}(k_{y}b)\right]
^{-1/2},  \label{pi}
\end{equation}%
where $R_{i}$ is the two-dimensional position of the Cu site $i$ and $b=a/2$%
, where $a$ is the lattice parameter of the CuO$_{2}$ planes. The new
operators $\pi _{i\sigma }$ satisfy canonical anticommutation rules. The
mapping is now different:

\begin{equation}
|i0\rangle \leftrightarrow |is\rangle =\frac{1}{\sqrt{2}}\left( \pi
_{i\uparrow }^{\dagger }d_{i\downarrow }^{\dagger }-\pi _{i\downarrow
}^{\dagger }d_{i\uparrow }^{\dagger }\right) |0\rangle .  \label{map}
\end{equation}

Inverting Eq. (\ref{pi}), one has for the two 2p$_{\sigma }$ O orbitals per
unit cell

\begin{eqnarray}
p_{i+b\mathbf{\hat{x}}\sigma } &=&\frac{1}{N}\sum_{\mathbf{k}}\beta _{%
\mathbf{k}}e^{-i\mathbf{k\cdot R}_{i}}e^{-ik_{x}b}\sum_{m}e^{i\mathbf{k\cdot
R}_{m}}\left[ \cos (k_{x}b)\pi _{m\sigma }+\cos (k_{y}b)\gamma _{m\sigma }%
\right] ,  \nonumber \\
p_{i+b\mathbf{\hat{y}}\sigma } &=&\frac{1}{N}\sum_{\mathbf{k}}\beta _{%
\mathbf{k}}e^{-i\mathbf{k\cdot R}_{i}}e^{-ik_{y}b}\sum_{m}e^{i\mathbf{k\cdot
R}_{m}}\left[ \cos (k_{y}b)\pi _{m\sigma }-\cos (k_{x}b)\gamma _{m\sigma }%
\right] ,  \label{pdelta}
\end{eqnarray}%
where the $\gamma _{m\sigma }$ correspond the so called non-bonding O
orbitals which do not mix with the Cu 3d$_{x^{2}-y^{2}}$ orbitals by
symmetry. They are defined asking that the Fourier transforms $\pi^\dagger
_{k\sigma }$ and $\gamma _{k\sigma }$ anticommute. In any case we neglect
these non-bonding orbitals in what follows.

Using Eqs. (\ref{pino}) and (\ref{pdelta}) one can write

\begin{eqnarray}
\tilde{\pi}_{i\sigma } &=&\sum_{m}\lambda (\mathbf{R}_{m})\pi _{i+m\sigma },
\label{pipi} \\
\lambda (\mathbf{R}_{m}) &=&\frac{1}{N}\sum_{\mathbf{k}}\left[ \cos
^{2}(k_{x}b)+\cos ^{2}(k_{y}b)\right] ^{1/2}\cos (\mathbf{k\cdot R}_{m}) =
\nonumber \\
&=&\frac{1}{N}\sum_{\mathbf{k}}\left[ 1+(\cos (k_{x}a)+\cos (k_{y}a))/2%
\right] ^{1/2}\cos (k_{x}x_{m})\cos (k_{y}y_{m}).  \label{lam}
\end{eqnarray}%
As expected, the sum in Eq. (\ref{pipi}) is dominated by $\lambda (\mathbf{R}%
_{m})\simeq 0.96$ and the other terms decrease rapidly with distance (see
Table \ref{integ}).

The part independent of spin of the first term in Eq. (\ref{hsf}) is

\begin{equation}
\frac{1}{2}(t_{1}^{sf}-t_{2}^{sf})\sum_{i\delta \delta ^{\prime }\sigma
}p_{i+\delta ^{\prime }\sigma }^{\dagger }p_{i+\delta \sigma
}=2(t_{1}^{sf}-t_{2}^{sf})\sum_{i\sigma }\tilde{\pi}_{i\sigma }^{\dagger }%
\tilde{\pi}_{i\sigma }=2(t_{1}^{sf}-t_{2}^{sf})\sum_{il\sigma }\nu (\mathbf{R%
}_{l})\pi _{i+l\sigma }^{\dagger }\pi _{i\sigma },  \label{prim}
\end{equation}%
where using Eqs. (\ref{pipi}), (\ref{lam}) and symmetry

\begin{eqnarray}
\nu (\mathbf{R}_{l}) &=&\sum_{m}\lambda (\mathbf{R}_{l}+\mathbf{R}%
_{m})\lambda (-\mathbf{R}_{m})=\frac{1}{N^{2}}\sum_{\mathbf{kq}m}(\beta _{%
\mathbf{k}}\beta _{\mathbf{q}})^{-1}e^{-i\mathbf{k\cdot (R}_{l}+\mathbf{R}%
_{m})}e^{i\mathbf{q\cdot R}_{m}} = \nonumber \\
&=&\frac{1}{N}\sum_{\mathbf{k}}(\beta _{\mathbf{k}})^{-2}e^{-i\mathbf{k\cdot
R}_{l}}=\frac{1}{N}\sum_{\mathbf{k}}\left[ 1+(\cos (k_{x}a)+\cos (k_{y}a))/2%
\right] \cos (k_{x}x_{l})\cos (k_{y}y_{l}).  \label{nu}
\end{eqnarray}%
It is easy to see that $\nu (\mathbf{0})=1$ (contributing to a constant
energy of the $\pi $ orbitals which we drop), $\nu (a\mathbf{\hat{x}})=\nu (a%
\mathbf{\hat{y}})=1/4$, and other $\nu (\mathbf{R}_{l})=0$. Calculating the
matrix element $\langle B|\pi _{j\uparrow }^{\dagger }\pi _{i\uparrow
}|A\rangle =-1/2$, where $|A\rangle =$ $d_{j\downarrow }^{\dagger
}|is\rangle $ and $|B\rangle =d_{i\downarrow }^{\dagger }|js\rangle $, one
realizes that the mapping Eq. (\ref{map}) leads to

\begin{equation}
P\pi _{j\uparrow }^{\dagger }\pi _{i\uparrow }P\longleftrightarrow
-d_{i\downarrow }^{\dagger }d_{j\downarrow }/2,  \label{mapo}
\end{equation}%
for the corresponding operators, and the same interchanging spin up and
down, where $P$ is the projector on the low-energy subspace of Zhang-Rice
singlets (LESZRS). Thus, the spin independent part of the first term in Eq. (%
\ref{hsf}) provides a contribution 
\begin{equation}
-\frac{1}{4}(t_{1}^{sf}-t_{2}^{sf})\sum_{i\delta \sigma }d_{i+2\delta \sigma
}^{\dagger }d_{i\sigma }  \label{h1}
\end{equation}%
to the NN hopping of the one-band model.

The spin dependent part of first term in Eq. (\ref{hsf}) is

\begin{equation}
(t_{1}^{sf}+t_{2}^{sf})\sum_{i\delta \delta ^{\prime }ss^{\prime
}}p_{i+\delta ^{\prime }s^{\prime }}^{\dagger }p_{i+\delta s}\mathbf{\sigma }%
_{s^{\prime }s}\cdot \mathbf{S}_{i}=4\sum_{iss^{\prime }}\tilde{\pi}%
_{is^{\prime }}^{\dagger }\tilde{\pi}_{is}\mathbf{\sigma }_{s^{\prime
}s}\cdot \mathbf{S}_{i},  \label{prims}
\end{equation}%
where $\mathbf{\sigma }_{s^{\prime }s}$ are the matrix elements between
spins $s^{\prime }$ and $s$ of a vector constructed from the three Pauli
matrices. Replacing Eq. (\ref{pipi}) in Eq. (\ref{prims}) one obtains
several terms. Note that for at most one added hole in the system $%
\sum_{ss^{\prime }}\pi _{ls^{\prime }}^{\dagger }\pi _{ms}\mathbf{\sigma }%
_{s^{\prime }s}\cdot \mathbf{S}_{i}=2\sum_{s}\pi _{ls}^{\dagger }\pi _{ms}%
\mathbf{s}_{m}\cdot \mathbf{S}_{i}=2\mathbf{s}_{l}\cdot \mathbf{S}%
_{i}\sum_{s}\pi _{ls}^{\dagger }\pi _{ms},$ where $\mathbf{s}%
_{l}=\sum_{ss^{\prime }}\pi _{ls^{\prime }}^{\dagger }\pi _{ls}\mathbf{%
\sigma }_{s^{\prime }s}/2$ is the spin of the Wannier function $\pi $ at
site $l$. Then if either $i=l$ or $i=m$, projection of this term in the
LESZRS Eq. (\ref{map}) leads to $\mathbf{s}_{i}\cdot \mathbf{S}_{i}=-3/4$,
and this term reduces to a hopping. Using $\lambda (-\mathbf{R}_{m})=\lambda
(\mathbf{R}_{m})$\ and neglecting as before the on-site energy correction
one obtains for the sum of all terms of this form

\begin{equation}
6(t_{1}^{sf}+t_{2}^{sf})\sum_{im\sigma }\lambda (\mathbf{R}_{m})\lambda (%
\mathbf{0})d_{i+m\sigma }^{\dagger }d_{i\sigma }.  \label{h2}
\end{equation}%
For the rest of the terms one can use $s_{m}^{z}=-S_{m}^{z}$ in the LESZRS,
Eq. (\ref{mapo}) and the mappings

\begin{equation}
P\pi _{j\uparrow }^{\dagger }\pi _{i\downarrow }P\longleftrightarrow
d_{i\uparrow }^{\dagger }d_{j\downarrow }/2=d_{i\uparrow }^{\dagger
}d_{j\uparrow }S_{j}^{+}/2,\text{ }P\pi _{j\downarrow }^{\dagger }\pi
_{i\uparrow }P\longleftrightarrow d_{i\downarrow }^{\dagger }d_{j\uparrow
}/2=d_{i\downarrow }^{\dagger }d_{j\downarrow }S_{j}^{-}/2,  \label{mapo2}
\end{equation}
leading to the following three-site terms

\begin{equation}
4(t_{1}^{sf}+t_{2}^{sf})\sum_{l\neq i\neq m\sigma }\lambda (\mathbf{R}_{l}-%
\mathbf{R}_{i})\lambda (\mathbf{R}_{m}-\mathbf{R}_{i})d_{l\sigma }^{\dagger
}d_{m\sigma }\mathbf{S}_{i}\cdot \mathbf{S}_{m}.  \label{h3}
\end{equation}

Using Eqs. (\ref{pdelta}) \ and neglecting non-bonding states absent in the
LESZRS, the second term of Eq. (\ref{hsf}) becomes

\begin{equation}
-J_{d}\sum_{lmiss^{\prime }}\eta (\mathbf{R}_{l}-\mathbf{R}_{i},\mathbf{R}%
_{m}-\mathbf{R}_{i})\pi _{ls^{\prime }}^{\dagger }\pi _{ms}\mathbf{\sigma }%
_{s^{\prime }s}\cdot \mathbf{S}_{i},  \label{jd}
\end{equation}%
where 
\begin{eqnarray}
\eta (\mathbf{R}_{l},\mathbf{R}_{m}) &=&\sum\limits_{\alpha =x,y}[A_{\alpha
}(\mathbf{R}_{l})A_{\alpha }(\mathbf{R}_{m})+B_{\alpha }(\mathbf{R}%
_{l})B_{\alpha }(\mathbf{R}_{m})],  \nonumber \\
A_{\alpha }(\mathbf{R}_{l}) &=&\frac{1}{N}\sum_{\mathbf{k}}\beta _{\mathbf{k}%
}\cos (k_{x}x_{l})\cos (k_{y}y_{l})\cos ^{2}(k_{\alpha }b),  \nonumber \\
B_{x}(\mathbf{R}_{l}) &=&-\frac{1}{2N}\sum_{\mathbf{k}}\beta _{\mathbf{k}%
}\sin (k_{x}x_{l})\cos (k_{y}y_{l})\sin (k_{x}a),  \nonumber \\
B_{x}(\mathbf{R}_{l}) &=&-\frac{1}{2N}\sum_{\mathbf{k}}\beta _{\mathbf{k}%
}\cos (k_{x}x_{l})\sin (k_{y}y_{l})\sin (k_{y}a).  \label{eta}
\end{eqnarray}%
As before we can separate from the sum the terms with either $i=l$ or $i=m$,
for which we can use $\mathbf{s}_{i}\cdot \mathbf{S}_{i}=-3/4$ in the
LESZRS, and the rest. Using also Eqs. (\ref{mapo}), (\ref{mapo2}) and $%
A_{\alpha }(-\mathbf{R}_{l})=A_{\alpha }(\mathbf{R}_{l})$, $B_{\alpha }(-%
\mathbf{R}_{l})=-B_{\alpha }(\mathbf{R}_{l})$, \ one obtains

\begin{equation}
-\frac{3}{2}J_{d}\sum_{il\sigma }\eta (\mathbf{R}_{l},\mathbf{0}%
)d_{i+l\sigma }^{\dagger }d_{i\sigma }-J_{d}\sum_{l\neq i\neq m\sigma }\eta (%
\mathbf{R}_{l}-\mathbf{R}_{i},\mathbf{R}_{m}-\mathbf{R}_{i})d_{l\sigma
}^{\dagger }d_{m\sigma }\mathbf{S}_{i}\cdot \mathbf{S}_{m}.  \label{h4}
\end{equation}

Finally, using Eqs (\ref{pdelta}) and neglecting non-bonding states the term
in $t_{pp}$ of Eq. (\ref{hsf}) becomes

\begin{equation}
\frac{t_{pp}}{2}\sum_{il\sigma }\mu (\mathbf{R}_{l})d_{i+l\sigma }^{\dagger
}d_{i\sigma },  \label{h5}
\end{equation}%
where 
\begin{equation}
\mu (\mathbf{R}_{l})=\frac{8}{N}\sum_{\mathbf{k}}\beta _{\mathbf{k}}^{2}\cos
^{2}(k_{x}b)\cos ^{2}(k_{y}b)  \label{mu}.
\end{equation}

Including the Cu-Cu exchange term and adding Eqs. (\ref{h1}), (\ref{h2}), (%
\ref{h3}), (\ref{h4}), and (\ref{h5}), one realizes that $PH_{sf}^{p}P$ can
be mapped into the following generalized $t-J$ model:

\begin{eqnarray}
H_{tJ}^{p} &=&\frac{t_{2}^{sf}-t_{1}^{sf}}{4}\sum_{i\delta \sigma
}d_{i+2\delta \sigma }^{\dagger }d_{i\sigma }+\sum_{im\sigma }\left[
6(t_{1}^{sf}+t_{2}^{sf})\lambda (\mathbf{R}_{m})\lambda (\mathbf{0})-\frac{3%
}{2}J_{d}\eta (\mathbf{R}_{m},\mathbf{0})+\frac{t_{pp}}{2}\mu (\mathbf{R}%
_{m})\right] d_{i+m\sigma }^{\dagger }d_{i\sigma } + \nonumber \\
&&+\sum_{l\neq i\neq m\sigma }\left[ 4(t_{1}^{sf}+t_{2}^{sf})\lambda (%
\mathbf{R}_{l}-\mathbf{R}_{i})\lambda (\mathbf{R}_{m}-\mathbf{R}%
_{i})-J_{d}\eta (\mathbf{R}_{l}-\mathbf{R}_{i},\mathbf{R}_{m}-\mathbf{R}_{i})%
\right] d_{l\sigma }^{\dagger }d_{m\sigma }\mathbf{S}_{i}\cdot \mathbf{S}_{m}+
\nonumber \\
&&+\frac{J}{2}\sum_{i\delta }\mathbf{S}_{i}\cdot \mathbf{S}_{_{i+2\delta }}.
\label{hptJ}
\end{eqnarray}%
The main two-dimensional integrals that enter this expression are displayed
in Table \ref{integ}. Note that $\lambda (\mathbf{R}_{l})$ and $\mu (\mathbf{%
R}_{l})$ are symmetric under the operations of the point group $C_{4v}$,
while $A_{\alpha }(\mathbf{-R}_{l})=A_{\alpha }(\mathbf{R}_{l})$ and $%
B_{\alpha }(\mathbf{-R}_{l})=-B_{\alpha }(\mathbf{R}_{l})$. Some of these
integrals were given previously \cite{bel2}. There are small differences in
some $\mu (\mathbf{R}_{l})$. We believe that our results are more accurate.

\begin{table}[h]
\caption{Two-dimensional integrals that enter $H_{tJ}$. See Eqs. (\protect
\ref{hptJ}) and (\protect\ref{eta}).}
\label{integ}%
\begin{ruledtabular}
\begin{tabular}{lllllll}
$\mathbf{R}/a$ & $\lambda$ & $A_x$ & $A_y$ & $B_x$ & $B_y$ & $\mu$ \\ 
\hline
(0,0)  & 0.9581 & 0.4791 & 0.4791 &  0  & 0  & 1.4535 \\
(1,0)  & 0.1401 & 0.1989 & -0.05877 &  0.2802  & 0  & 0.5465 \\
(1,1)  & -0.02351 & -0.01753 & -0.01753 &    &   & 0.2441 \\
(2,0)  & -0.01373 & -0.02643 & 0.01270 &    &   & -0.1277 \\

\end{tabular}
\end{ruledtabular}
\end{table}

\section{The generalized $t-J$ model for T-CuO}
\label{gtjt}

Naturally, the one-band model for the other CuO$_{2}$ sublattice $H_{tJ}^{q}$
(the mapping of $PH_{sf}^{q}P$ to a generalized $t-J$ model) has the same
form as $H_{tJ}^{p}$ above, with the only difference that $i$ refers to Cu
sites of the other sublattice. In addition, the exchange term proportional
to $J^{\prime }$ in Eq. (\ref{h}) retains the same form in the one-band
model. Therefore, the remaining task is to map the term proportional to $%
t_{pp}^{\prime }$.

\subsection{Mapping using non-orthogonal singlets}
\label{no}

\begin{figure}[h]
\includegraphics[width=7. cm]{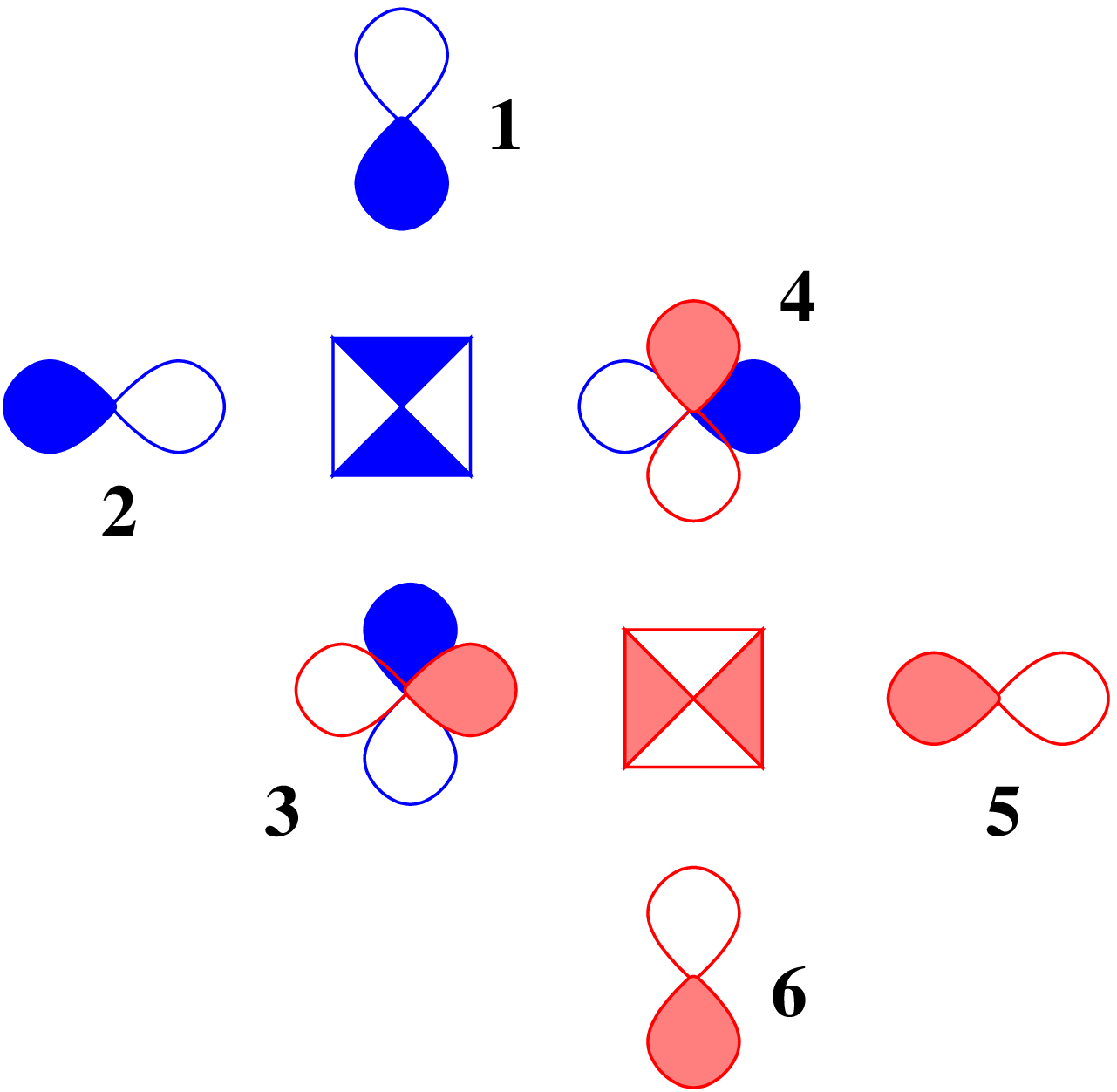}
\caption{Orbitals involved in the hopping between non-orthogonal ZRS
belonging to different CuO$_2$ sublattices.}
\label{t0}
\end{figure}
We define the nonorthogonal ZRS for the second CuO$_{2}$ sublattice in
analogy to Eqs. (\ref{zr}) and (\ref{pino}):

\begin{eqnarray}
|i\tilde{s}\rangle &=&\frac{1}{\sqrt{2}}\left( \tilde{\rho}_{i\uparrow
}^{\dagger }d_{i\downarrow }^{\dagger }-\tilde{\rho}_{i\downarrow }^{\dagger
}d_{i\uparrow }^{\dagger }\right) |0\rangle ,\text{ }  \label{zr2} \\
\text{ }\tilde{\rho}_{i\sigma } &=&\frac{1}{2}\sum_{\delta }q_{i+\delta
\sigma },  \label{rhono}
\end{eqnarray}%
Adolphs \textit{et al}. argue that the orbitals $\tilde{\pi}_{i\sigma }$ and 
$\tilde{\rho}_{n\sigma }$ at NN sites do not mix \cite{adol}. However, in
spite of a partial cancellation, the result is nonzero. An example is shown
in Fig. \ref{t0} for $\mathbf{R}_{n}=\mathbf{R}_{i}+b(\mathbf{\hat{x}}-\mathbf{\hat{%
y}})$. In terms of the numbers of the figure

\begin{equation}
\tilde{\pi}_{i\sigma }=\frac{1}{2}\sum_{i=1}^{4}p_{i\sigma },\text{ }\tilde{%
\rho}_{n\sigma }=\sum_{i=3}^{6}q_{i\sigma }.  \label{zrno2}
\end{equation}%
Then

\begin{equation}
t_{pp}^{\prime }\sum_{ij\gamma \sigma }s_{\gamma }\left( p_{j+\gamma \sigma
}^{\dagger }q_{j\sigma }+\mathrm{H.c.}\right) \tilde{\pi}_{i\sigma
}^{\dagger }|0\rangle =\frac{t_{pp}^{\prime }}{2}(q_{5\sigma }^{\dagger
}+q_{5\sigma }^{\dagger })|0\rangle +...=\frac{t_{pp}^{\prime }}{2}\tilde{%
\rho}_{n\sigma }^{\dagger }|0\rangle +...  \label{mixno}
\end{equation}%
It is easy to see that the same value $t_{pp}^{\prime }/2$ is obtained for $%
\mathbf{R}_{n}-\mathbf{R}_{i}=-b(\mathbf{\hat{x}}-\mathbf{\hat{y}})$, while
the result is $-t_{pp}^{\prime }/2$ for $\mathbf{R}_{n}-\mathbf{R}_{i}=\pm b(%
\mathbf{\hat{x}}+\mathbf{\hat{y}})$. There are also contributions $\pm
t_{pp}^{\prime }/4$ at fourth NN. The mapping $|i\tilde{s}\rangle
\leftrightarrow |i0\rangle $ leads to a factor -1/2 [similar to Eq. (\ref{mapo})]
plus some corrections due to non-orthogonality of the ZRS \cite{sys}. 
The details are beyond the scope of this work. In the following
subsection, we derive the rigorous result using orthogonal ZRS. In any case,
the simpler results presented here show that the effective hopping is not
zero.

\subsection{Mapping using orthogonal singlets}
\label{os}

The term in $t_{pp}^{\prime }$ of Eq. (\ref{h}) can be written in the form

\begin{eqnarray}
H^{\prime } &=&t_{pp}^{\prime }\sum_{ij\gamma \sigma }s_{\gamma }\left(
p_{j+\gamma \sigma }^{\dagger }q_{j\sigma }+{\rm H.c.}\right) =  \nonumber
\\
&=&t_{pp}^{\prime }\sum_{i\sigma }[p_{i+b\mathbf{\hat{x}}\sigma }^{\dagger
}(q_{i+b\mathbf{\hat{y}}\sigma }+q_{i+a\mathbf{\hat{x}-}b\mathbf{\hat{y}}%
\sigma }-q_{i-b\mathbf{\hat{y}}\sigma }-q_{i+a\mathbf{\hat{x}-}b\mathbf{\hat{%
y}}\sigma })+  \nonumber \\
&&+p_{i+b\mathbf{\hat{y}}\sigma }^{\dagger }(q_{i+b\mathbf{\hat{x}}\sigma
}+q_{i-b\mathbf{\hat{x}+}a\mathbf{\hat{y}}\sigma }-q_{i-b\mathbf{\hat{x}}%
\sigma }+q_{i+b\mathbf{\hat{x}+}a\mathbf{\hat{y}}\sigma })+\mathrm{H.c.}],
\label{tpp}
\end{eqnarray}%
where the sum runs over all sites of the first CuO$_{2}$ sublattice.

Using Eqs. (\ref{pdelta}) and the corresponding ones for the second CuO$_{2}$
sublattice:

\begin{eqnarray}
q_{n+b\mathbf{\hat{x}}\sigma } &=&\frac{1}{N}\sum_{\mathbf{k}}\beta _{%
\mathbf{k}}e^{-i\mathbf{k\cdot R}_{n}}e^{-ik_{x}b}\sum_{m}e^{i\mathbf{k\cdot
R}_{m}}\left[ \cos (k_{x}b)\rho _{m\sigma }+...\right] ,  \nonumber \\
q_{n+b\mathbf{\hat{y}}\sigma } &=&\frac{1}{N}\sum_{\mathbf{k}}\beta _{%
\mathbf{k}}e^{-i\mathbf{k\cdot R}_{n}}e^{-ik_{y}b}\sum_{m}e^{i\mathbf{k\cdot
R}_{m}}\left[ \cos (k_{y}b)\rho _{m\sigma }-...\right] ,  \label{qdelta}
\end{eqnarray}%
one obtains after some algebra

\begin{equation}
H^{\prime }=t_{pp}^{\prime }\xi (\mathbf{R}_{\tau })\sum_{i\tau \sigma
}\left( \pi _{i\sigma }^{\dagger }\rho _{i+\tau \sigma }+\mathrm{H.c.}%
\right) ,  \label{tpp2}
\end{equation}%
where $\tau $ denotes the vectors connecting both CuO$_{2}$ sublattices ($%
x_{\tau }$ and $y_{\tau }$ below are both odd multiples of $b$) and

\begin{equation}
\xi (\mathbf{R}_{\tau })=-\frac{4}{N}\sum\limits_{k}\sin (k_{x}b)\sin
(k_{y}b)\sin (k_{x}x_{\tau })\sin (k_{y}y_{\tau }).  \label{xi}
\end{equation}%
It is easy to see that $\xi (\mathbf{R}_{\tau })=-1$ if $\mathbf{R}_{\tau
}=\pm b(\mathbf{\hat{x}}+\mathbf{\hat{y}})$, $\xi (\mathbf{R}_{\tau })=1$ if 
$\mathbf{R}_{\tau }=\pm b(\mathbf{\hat{x}}-\mathbf{\hat{y}})$, and $\xi (%
\mathbf{R}_{\tau })=0$ for other $\mathbf{R}_{\tau }$. Therefore

\begin{equation}
H^{\prime }=t_{pp}^{\prime }\sum_{i\gamma \sigma }s_{\gamma }\left( \pi
_{i\sigma }^{\dagger }\rho _{i+\gamma \sigma }+\mathrm{H.c.}\right) .
\label{tpp3}
\end{equation}%
using the mapping Eq. (\ref{mapo}) and adding the other terms, the complete
generalized $t-J$ model for T-CuO takes the form

\begin{equation}
H_{tJ}=H_{tJ}^{p}+H_{tJ}^{q}-\frac{t_{pp}^{\prime }}{2}\sum_{i\gamma \sigma
}s_{\gamma }\left( d_{i\sigma }^{\dagger }d_{i+\gamma \sigma }+\mathrm{H.c.}%
\right) -\frac{J^{\prime }}{2}\sum_{i\gamma }\mathbf{S}_{i}\cdot \mathbf{S}%
_{_{i+\gamma }}.  \label{htj}
\end{equation}

To compare with experiment it is convenient to write the Hamiltonian in
terms of the following operators

\begin{equation}
c_{i\sigma }=e^{i\mathbf{Q}\cdot (\mathbf{R}_{i}-\mathbf{R}%
_{i}^{0})}d_{i\sigma },\text{ }  \label{cd}
\end{equation}%
which restores the original phases of the Cu orbitals [changed before in Eq.
(\ref{phases})]. If the phases are not restored, the problem is of course
equivalent, but the wave vectors are displaced by $\mathbf{Q}$ complicating
the comparison with experiment. This transformation within each CuO$_{2}$
sublattice changes the sign of the NN hopping (at distances $\pm a\mathbf{%
\hat{x}}$, $\pm a\mathbf{\hat{y}}$) leaving second and third NN hopping
unchanged. In addition also the sign of the intersublattice hopping at
distances $\pm b(\mathbf{\hat{x}}-\mathbf{\hat{y}})$ is changed, keeping the
sign in the perpendicular direction, so that the corresponding term in Eq. (%
\ref{htj}) becomes

\begin{equation}
H_{NN}=\frac{t_{pp}^{\prime }}{2}\sum_{i\gamma \sigma }\left( c_{i\sigma
}^{\dagger }c_{i+\gamma \sigma }+\mathrm{H.c.}\right) .  \label{hnn}
\end{equation}

\section{Simplified generalized $t-J$ model}
\label{sgtj}

The state of the art technique for studying the dynamics of one hole in an
antiferromagnet is the self-consistent Born approximation (SCBA) \cite{mh,lema,lema2,trum}. 
It compares very well with exact diagonalization of
small clusters \cite{mh,lema2,trum, Hamad08}, while permitting an extensions to larger
clusters. From previous studies for the antiferromagnetic order of CuO$_{2}$
planes, one knows that the propagation of the hole is easier through each
sublattice with spins pointing in the same direction, in particular for
hopping involving second and third NN, while it is inhibited for first NN in
spite of the fact the corresponding hopping is larger, because the hopping
distorts the antiferromagnetic alignement.

The generalized $t-J$ model for CuO$_{2}$ planes, as described above,
contains three-site terms which combine second and third NN with spin-flip
processes. These so called correlated hopping processes are argued to play
an important role for superconductivity \cite{rvb,supco}. However, the above
argument indicates that for the propagation of the hole, only the
spin-conserving part is important. Therefore, to simplify the model and
bring it amenable to the SCBA treatment we retain only hopping up to third
NN in the CuO$_{2}$ planes and approximate $\mathbf{S}_{i}\cdot \mathbf{S}%
_{m}\simeq \langle S_{i}^{z}S_{m}^{z}\rangle $ in Eq. (\ref{hptJ}). This
leads to a simplified effective model for T-CuO similar to that considered
by Moser \textit{et al} \cite{moser}.

\begin{equation}
H_{tJ}^{s}=-\sum_{\kappa =0}^{3}t_{\kappa }\sum_{iv_{\kappa }\sigma }\left(
c_{i\sigma }^{\dagger }c_{i+v_{\kappa }\sigma }+\mathrm{H.c.}\right) +\frac{J%
}{2}\sum_{iv_{1}}\mathbf{S}_{i}\cdot \mathbf{S}_{_{i+v_{1}}}-\frac{J^{\prime
}}{2}\sum_{iv_{0}}\mathbf{S}_{i}\cdot \mathbf{S}_{_{i+v_{0}}},  \label{hstj}
\end{equation}%
where the subscript $\kappa =0$ refers to intersublattice hopping of NN Cu
atoms in the T-CuO structure (connected by the vectors $v_{0}=\pm b(\mathbf{%
\hat{x}\pm \hat{y}})$), while $\kappa =1$, 2, 3, refer to first, second, and
third NN within each CuO$_{2}$ sublattice. Comparison with Eqs. (\ref{hptJ}%
), (\ref{hnn}) and using Eq. (\ref{cd}) leads to

\begin{eqnarray}
t_{0} &=&-\frac{t_{pp}^{\prime }}{2},  \nonumber \\
t_{1} &=&\frac{t_{2}^{sf}-t_{1}^{sf}}{4}+6(t_{1}^{sf}+t_{2}^{sf})\lambda (%
\mathbf{0})\lambda (a\mathbf{\hat{x}})-\frac{3}{2}J_{d}\eta (a\mathbf{\hat{x}%
},\mathbf{0})+\frac{t_{pp}}{2}\mu (a\mathbf{\hat{x}}),  \nonumber \\
-t_{2} &\simeq &6(t_{1}^{sf}+t_{2}^{sf})\lambda (\mathbf{0})\lambda \left( a(%
\mathbf{\hat{x}+\hat{y})}\right) -\frac{3}{2}J_{d}\eta \left( a(\mathbf{\hat{%
x}+\hat{y})},\mathbf{0}\right) +\frac{t_{pp}}{2}\mu \left( a(\mathbf{\hat{x}+%
\hat{y})}\right) + \nonumber \\
&&+2\langle S_{i}^{z}S_{i+v_{1}}^{z}\rangle \left[ 4(t_{1}^{sf}+t_{2}^{sf})%
\lambda ^{2}(a\mathbf{\hat{x}})-J_{d}\eta (a\mathbf{\hat{x}},a\mathbf{\hat{y}%
})\right] ,  \nonumber \\
-t_{3} &\simeq &6(t_{1}^{sf}+t_{2}^{sf})\lambda (\mathbf{0})\lambda \left( 2a%
\mathbf{\hat{x}}\right) -\frac{3}{2}J_{d}\eta \left( 2a\mathbf{\hat{x}},%
\mathbf{0}\right) +\frac{t_{pp}}{2}\mu \left( 2a\mathbf{\hat{x}}\right) 
\nonumber + \\
&&+\langle S_{i}^{z}S_{i+v_{1}}^{z}\rangle \left[ 4(t_{1}^{sf}+t_{2}^{sf})%
\lambda ^{2}(a\mathbf{\hat{x}})-J_{d}\eta (a\mathbf{\hat{x}},-a\mathbf{\hat{x%
}})\right] .  \label{ts}
\end{eqnarray}%
Using Eqs. (\ref{eta}), Table \ref{integ}, and $\langle
S_{i}^{z}S_{i+v_{1}}^{z}\rangle =-0.186$ for the NN expectation value for
the Heisenberg model in the square lattice (see next section \ref{ssc} ), one obtains

\begin{eqnarray*}
t_{1} &\simeq &0.555t_{1}^{sf}+1.055t_{2}^{sf}+0.273t_{pp}-0.101J_{d}, \\
t_{2} &\simeq &0.161(t_{1}^{sf}+t_{2}^{sf})-0.122t_{pp}-0.0173J_{d}, \\
t_{3} &\simeq &0.0935(t_{1}^{sf}+t_{2}^{sf})+0.0638t_{pp}-0.0033J_{d}.
\end{eqnarray*}%
The fact that $t_{1}^{sf}+t_{2}^{sf}$ and $t_{pp}$ enter with different sign
in $t_{2}$ leads to a large relative error in this parameter. Fortunately,
the results seem to be rather insensitive to $t_{2}$.

Using the estimated parameters for the spin-fermion model based on previous
constrained-density-functional calculations (set A) or given by Adolphs 
\textit{et al} \cite{adol} (set B),
tabulated in Table \ref{parsf}, we obtain the results presented in Table \ref{pareff}.

\begin{table}[h]
\caption{Parameters of the effective model for T-CuO in meV.}
\label{pareff}%
\begin{ruledtabular}
\begin{tabular}{lllllll}
set & $t_{0}$ & $t_{1}$ & $t_{2}$ & $t_{3}$  & $J$ & $J^{\prime}$ \\ 
\hline
A & -168 & 417 &   -2 & 69 & 130 & 3 \\
B & -184 & 369 & -11 & 65 & 150 & 0 \\

\end{tabular}
\end{ruledtabular}
\end{table}

\section{The self-consistent Born approximation}
\label{scba}

As Adolphs \textit{et al}. \cite{adol}, we assume the antiferromagnetic order
of T-CuO given in the left of Fig. \ref{ordenesmagneticos}. The NN Cu atoms connected by 
the vectors $\pm \mathbf{c}$ ($\pm \mathbf{d}$), have parallel (antiparallel) spins, 
where $\mathbf{c}=b(\mathbf{\hat{x}}+\mathbf{\hat{y}})$ and $\mathbf{d}=b(-\mathbf{\hat{x}}+%
\mathbf{\hat{y}})$. The primitive translation vectors in the plane, which
also define the unit cell, are $\mathbf{c}$ and $2\mathbf{d}$. The unit cell
has the same size as that of the CuO$_{2}$ planes but it is different.

\begin{figure*}
{\includegraphics[width=0.3\columnwidth]{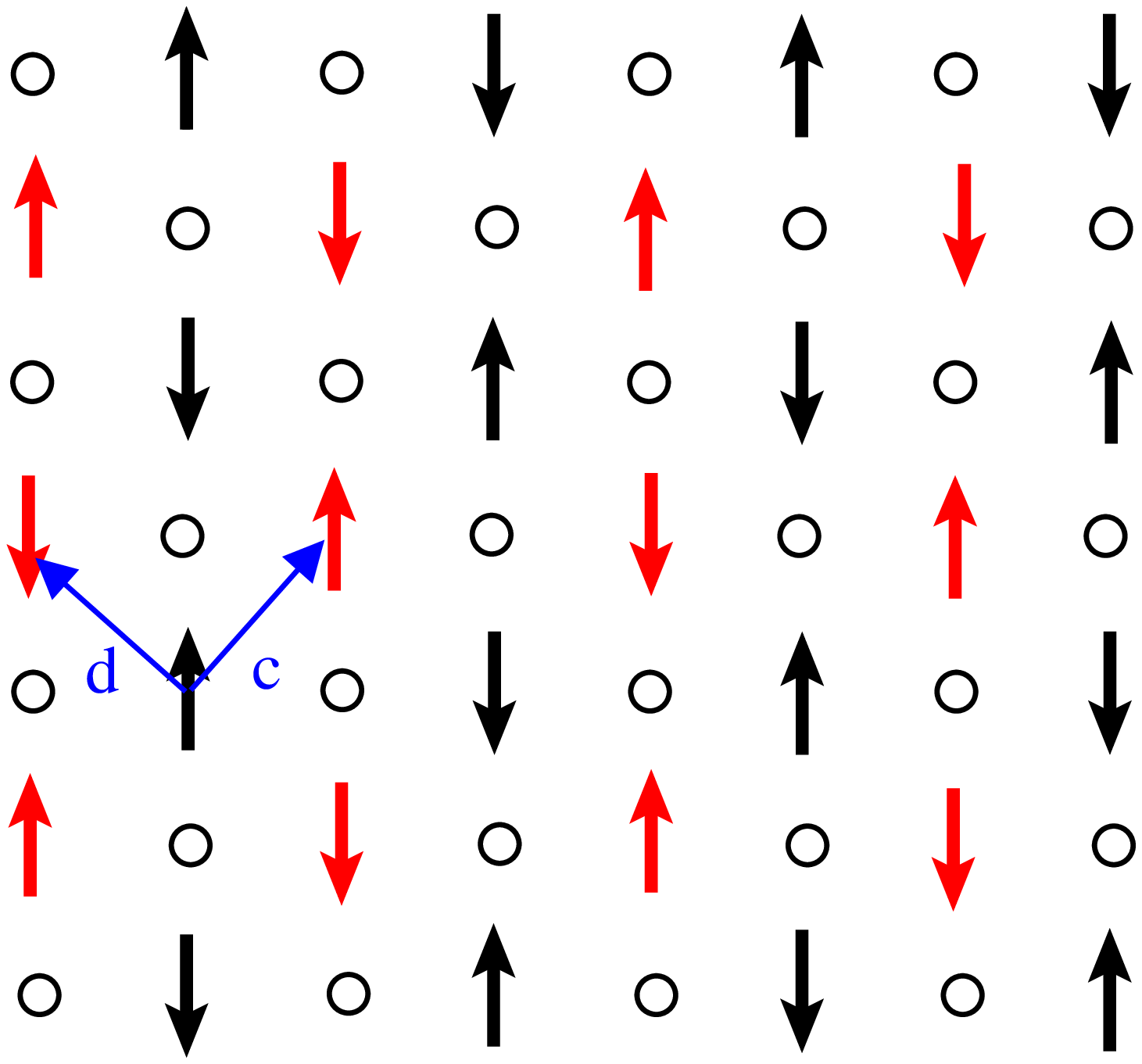}} \hspace{2cm}
{\includegraphics[width=0.3\columnwidth]{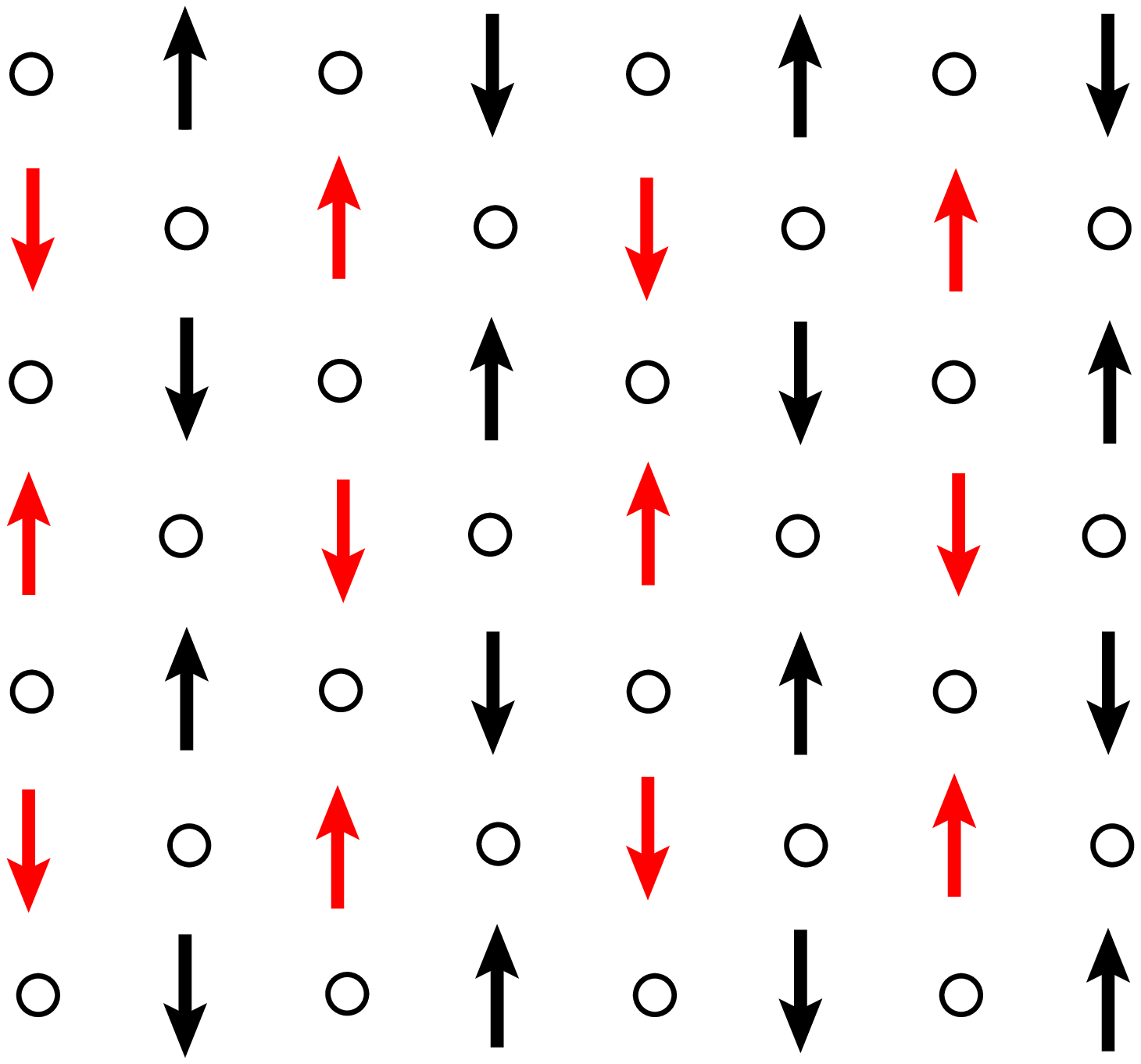}}
\caption{(Color online) The two possible magnetic ground states for T-CuO 
$\mathbf{Q}=(0,\pi)$ (left) and $\mathbf{Q}=(\pi,0)$ (right). 
The vectors $\bf{c}$ and $\bf{d}$ are indicated in the left figure.}
\label{ordenesmagneticos}
\end{figure*}

Following Mart\'{\i}nez and Horsch \cite{mh}, we perform the transformation $%
c_{i\sigma }\longrightarrow c_{i-\sigma }$ in the sublattice in which the
spins are pointing down \cite{note2}, in such a way that all spins are pointing
up after the transformation. Then we define the spin excitations $%
a_{i}^{\dagger }=c_{i\downarrow }^{\dagger }c_{i\uparrow }$, and the holon
operators $h_{i}$ such that

\begin{equation}
c_{i\uparrow }=h_{i}^{\dagger }\text{, }c_{i\downarrow }=h_{i}^{\dagger
}a_{i}\text{,}  \label{holon}
\end{equation}%
in the Hilbert subspace we are considering.

\subsection{Spin waves}
\label{sw}

We first diagonalize the exchange part of the Hamiltonian Eq. (\ref{hstj})
for the undoped system. In terms of the spin excitations, it takes the form

\begin{eqnarray}
H_{e} &=&\frac{J}{2}\sum_{iv_{1}}\mathbf{S}_{i}\cdot \mathbf{S}_{_{i+v_{1}}}-%
\frac{J^{\prime }}{2}\sum_{iv_{0}}\mathbf{S}_{i}\cdot \mathbf{S}%
_{_{i+v_{0}}}=-2JN + \nonumber \\
&&+\frac{J}{4}\sum_{iv_{1}}\left( a_{i}^{\dagger }a_{i}+a_{i+v_{1}}^{\dagger
}a_{i+v_{1}}+a_{i}a_{i+v_{1}}+a_{i}^{\dagger }a_{i+v_{1}}^{\dagger }\right) -
\nonumber \\
&&-\frac{J^{\prime }}{2}\sum_{i}\left( a_{i}^{\dagger }a_{i}+a_{i+\mathbf{d}%
}^{\dagger }a_{i+\mathbf{d}}+a_{i}a_{i+\mathbf{d}}+a_{i}^{\dagger }a_{i+%
\mathbf{d}}^{\dagger }\right)   \nonumber +\\
&&+\frac{J^{\prime }}{2}\sum_{i}\left( a_{i}^{\dagger }a_{i}+a_{i+\mathbf{c}%
}^{\dagger }a_{i+\mathbf{c}}-a_{i}^{\dagger }a_{i+\mathbf{c}}-a_{i+\mathbf{c}%
}^{\dagger }a_{i}\right) .  \label{he}
\end{eqnarray}%
Using the Fourier transform $a_{i}=N^{-1/2}\sum_{\mathbf{k}}e^{-i\mathbf{%
k\cdot R}_{i}}a_{\mathbf{k}}$, one obtains

\begin{eqnarray}
H_{e}+2JN &=&\sum_{\mathbf{k}}\left[ A_{\mathbf{k}}a_{\mathbf{k}}^{\dagger
}a_{\mathbf{k}}+B_{\mathbf{k}}\left( a_{\mathbf{k}}a_{-\mathbf{k}}+\mathrm{%
H.c.}\right) \right] ,  \nonumber \\
A_{\mathbf{k}} &=&2J-J^{\prime }\cos (\mathbf{c}\cdot \mathbf{k})\text{, }B_{%
\mathbf{k}}=\frac{J}{4}\sum_{v_{1}}\cos (v_{1}\cdot \mathbf{k})-\frac{%
J^{\prime }}{2}\cos (\mathbf{d}\cdot \mathbf{k}).  \label{he2}
\end{eqnarray}%
Dropping the constant $2JN$, $H_{e}$ is set into diagonal form
introducing new bosonic operators:

\begin{eqnarray}
H_{e} &=&\sum_{\mathbf{k}}\omega _{\mathbf{k}}\theta _{\mathbf{k}}^{\dagger
}\theta _{\mathbf{k}},\text{ }\theta _{\mathbf{k}}=u_{\mathbf{k}}a_{\mathbf{k%
}}+v_{\mathbf{k}}a_{-\mathbf{k}}^{\dagger },  \nonumber \\
\omega _{\mathbf{k}} &=&\sqrt{A_{\mathbf{k}}^{2}-4B_{\mathbf{k}}^{2},\text{ }%
}u_{\mathbf{k}}^{2}=\frac{1}{2}+\frac{A_{\mathbf{k}}}{2\omega _{\mathbf{k}}},%
\text{ }v_{\mathbf{k}}^{2}=u_{\mathbf{k}}^{2}-1,  \nonumber \\
u_{\mathbf{k}} &>&0,\text{sgn}(v_{\mathbf{k}})=\text{sgn}(B_{\mathbf{k}}).
\label{magn}
\end{eqnarray}

\subsubsection{Spin-spin correlations}
\label{ssc}

In the spin-wave formalism, the correlation function entering Eq. (\ref{ts})
becomes

\begin{equation}
\langle S_{i}^{z}S_{i+v_{1}}^{z}\rangle =\langle \left( \frac{1}{2}%
-a_{i}^{\dagger }a_{i}\right) \left( -\frac{1}{2}+a_{i+v_{1}}^{\dagger
}a_{i+v_{1}}\right) \rangle =-\frac{1}{4}+\langle a_{i}^{\dagger
}a_{i}\rangle -\langle a_{i}^{\dagger }a_{i}a_{i+v_{1}}^{\dagger
}a_{i+v_{1}}\rangle ,  \label{szsz}
\end{equation}%
where we have taken into account that the spins of sites $i$ and $i+v_{1}$
point in opposite directions. Decoupling the last correlation
function%
\begin{equation}
\langle a_{i}^{\dagger }a_{i}a_{i+v_{1}}^{\dagger }a_{i+v_{1}}\rangle
=\langle a_{i}^{\dagger }a_{i}\rangle \langle a_{i+v_{1}}^{\dagger
}a_{i+v_{1}}\rangle +|\langle a_{i}^{\dagger }a_{i+v_{1}}^{\dagger }\rangle
|^{2}+|\langle a_{i}^{\dagger }a_{i+v_{1}}\rangle |^{2},  \label{cf}
\end{equation}%
we obtain

\begin{equation}
\langle S_{i}^{z}S_{i+v_{1}}^{z}\rangle =-m^{2}-|\langle a_{i}^{\dagger
}a_{i+v_{1}}^{\dagger }\rangle |^{2}-|\langle a_{i}^{\dagger
}a_{i+v_{1}}\rangle |^{2},  \label{szsz2}
\end{equation}%
where 

\begin{equation}
m=\frac{1}{2}-\langle a_{i}^{\dagger }a_{i}\rangle   \label{m}
\end{equation}%
is the sublattice magnetization.

Transforming Fourier and suing the inverse of the second Eq. (\ref{magn}) 

\begin{equation}
a_{\mathbf{k}}=u_{\mathbf{k}}\theta _{\mathbf{k}}-v_{\mathbf{k}}\theta _{-%
\mathbf{k}}^{\dagger },  \label{theta}
\end{equation}%
the different correlation functions become at zero temperature

\begin{eqnarray}
\langle a_{i}^{\dagger }a_{i}\rangle  &=&\frac{1}{N}\sum_{\mathbf{kq}%
}\langle \left( u_{\mathbf{k}}\theta _{\mathbf{k}}^{\dagger }-v_{\mathbf{k}%
}\theta _{-\mathbf{k}}\right) \left( u_{\mathbf{q}}\theta _{\mathbf{q}}-v_{%
\mathbf{q}}\theta _{-\mathbf{q}}^{\dagger }\right) \rangle =\frac{1}{N}\sum_{%
\mathbf{k}}v_{\mathbf{k}}^{2},  \nonumber \\
\langle a_{i}^{\dagger }a_{i+v_{1}}^{\dagger }\rangle  &=&\frac{1}{N}\sum_{%
\mathbf{kq}}\langle \left( u_{\mathbf{k}}\theta _{\mathbf{k}}^{\dagger }-v_{%
\mathbf{k}}\theta _{-\mathbf{k}}\right) e^{iq\mathbf{\cdot }v_{1}}\left( u_{%
\mathbf{q}}\theta _{\mathbf{q}}^{\dagger }-v_{\mathbf{q}}\theta _{-\mathbf{q}%
}\right) \rangle =\frac{1}{N}\sum_{\mathbf{k}}\cos (\mathbf{k}\cdot v_{1})u_{%
\mathbf{q}}v_{\mathbf{k}},  \nonumber \\
\langle a_{i}^{\dagger }a_{i+v_{1}}\rangle  &=&\frac{1}{N}\sum_{\mathbf{kq}%
}\langle \left( u_{\mathbf{k}}\theta _{\mathbf{k}}^{\dagger }-v_{\mathbf{k}%
}\theta _{-\mathbf{k}}\right) e^{iq\mathbf{\cdot }v_{1}}\left( u_{\mathbf{q}%
}\theta _{\mathbf{q}}-v_{\mathbf{q}}\theta _{-\mathbf{q}}^{\dagger }\right)
\rangle =\frac{1}{N}\sum_{\mathbf{k}}\cos (\mathbf{k}\cdot v_{1})v_{\mathbf{k%
}}^{2},  \label{mv}
\end{eqnarray}%
We have evaluated the two-dimensional integrals above for $J^{\prime }=0$.
The result is $\langle a_{i}^{\dagger }a_{i}\rangle =0.19660$,  $\langle
a_{i}^{\dagger }a_{i+v_{1}}^{\dagger }\rangle =0.27558$ and $\langle
a_{i}^{\dagger }a_{i+v_{1}}\rangle =0$, leading to $m=-0.30340$ and $\langle
S_{i}^{z}S_{i+v_{1}}^{z}\rangle =-0.16799.$

\subsection{The hopping terms}
\label{hop}

The hopping terms of the Hamiltonian Eq. (\ref{hstj}) can be separated in
two: those involving two sites of the same sublattice (spin up or down),
like the terms in $t_{2}$ and $t_{3}$, and those connecting sites of
different sublattices ($t_{1}$ and half of the terms in $t_{0}$). The latter
give rise to a holon-magnon interaction. We neglect the terms creating two
spin excitations. Using the transformations introduced at the beginning of
this section we obtain 

\begin{eqnarray}
H_{t} &=&-\sum_{\kappa =0}^{3}t_{\kappa }\sum_{iv_{\kappa }\sigma }\left(
c_{i\sigma }^{\dagger }c_{i+v_{\kappa }\sigma }+\mathrm{H.c.}\right)
=t_{0}\sum_{i}\left[ h_{i}^{\dagger }h_{i+\mathbf{c}}+h_{i}^{\dagger }h_{i+%
\mathbf{d}}\left( a_{i}+a_{i+\mathbf{d}}\right) +\mathrm{H.c.}\right]  +
\nonumber \\
&&+t_{0}\sum_{i}\left( h_{i}^{\dagger }a_{i}\sum_{v_{1}}h_{i+v_{1}}+\mathrm{%
H.c.}\right) +\sum_{\kappa =2}^{3}t_{\kappa }\sum_{iv_{\kappa }\sigma
}\left( h_{i\sigma }^{\dagger }h_{i+v_{\kappa }\sigma }+\mathrm{H.c.}\right)
.  \label{ht}
\end{eqnarray}%
Using $h_{i}=N^{-1/2}\sum_{\mathbf{k}}e^{-i\mathbf{k\cdot R}_{i}}h_{\mathbf{k%
}}$, Eq. (\ref{theta}), and adding $H_{e}=\sum_{\mathbf{k}}\omega _{\mathbf{k}%
}\theta _{\mathbf{k}}^{\dagger }\theta _{\mathbf{k}}$ we obtain, after some
algebra

\begin{eqnarray}
H_{tJ}^{s} &=&\sum_{\mathbf{k}}\epsilon _{\mathbf{k}}h_{\mathbf{k}}^{\dagger
}h_{\mathbf{k}}+\sum_{\mathbf{k}}\omega _{\mathbf{k}}\theta _{\mathbf{k}%
}^{\dagger }\theta _{\mathbf{k}}+\frac{1}{\sqrt{N}}\left( \sum_{\mathbf{kq}%
}M_{\mathbf{kq}}h_{\mathbf{k}}^{\dagger }h_{\mathbf{k}-\mathbf{q}}\theta _{%
\mathbf{q}}+\mathrm{H.c.}\right) ,  \nonumber \\
\epsilon _{\mathbf{k}} &=&2t_{0}\cos (\mathbf{k}\cdot \mathbf{c})+4t_{2}\cos
(ak_{x})\cos (ak_{y})+2t_{3}\left[ \cos (2ak_{x})+\cos (2ak_{y})\right] , 
\nonumber \\
M_{\mathbf{kq}} &=&2t_{0}\left\{ \cos \left[ (\mathbf{k-q})\cdot \mathbf{c}%
\right] u_{\mathbf{q}}-\cos (\mathbf{k}\cdot \mathbf{c})v_{\mathbf{q}%
}\right\} +2t_{1}\left[ u_{\mathbf{q}}\zeta (\mathbf{k-q})-v_{\mathbf{q}%
}\zeta (\mathbf{k})\right] ,  \nonumber \\
\zeta (\mathbf{k}) &=&\cos (ak_{x})+\cos (ak_{y}).  \label{hfinal}
\end{eqnarray}

The holon Green function $G_{h}(\mathbf{k},\omega $) is obtained from the
self-consistent solution of the following equations:

\begin{eqnarray}
G_{h}^{-1}(\mathbf{k},\omega ) &=&\omega -\epsilon _{\mathbf{k}}-\Sigma (%
\mathbf{k},\omega )+i\epsilon ,  \nonumber \\
\Sigma (\mathbf{k},\omega ) &=&\frac{1}{N}\sum_{\mathbf{q}}M_{\mathbf{kq}%
}^{2}G_{h}(\mathbf{k}-\mathbf{q},\omega -\omega _{\mathbf{q}}).
\label{escba}
\end{eqnarray}%
In practice, the calculations are done in a large but finite system and the
selfconsistency can be avoided calculating sequentially 
$\Sigma (\mathbf{k},\omega )$ for increasing values of $\omega $, beginning with values (near $-4J$) such
that $\Sigma (\mathbf{k},\omega -\omega _{\mathbf{q}})=0$ for all $\mathbf{k}
$ and $\mathbf{q}$ \cite{HamadThesis}.  

\begin{figure*}[t]
{\includegraphics[width=0.5\columnwidth]{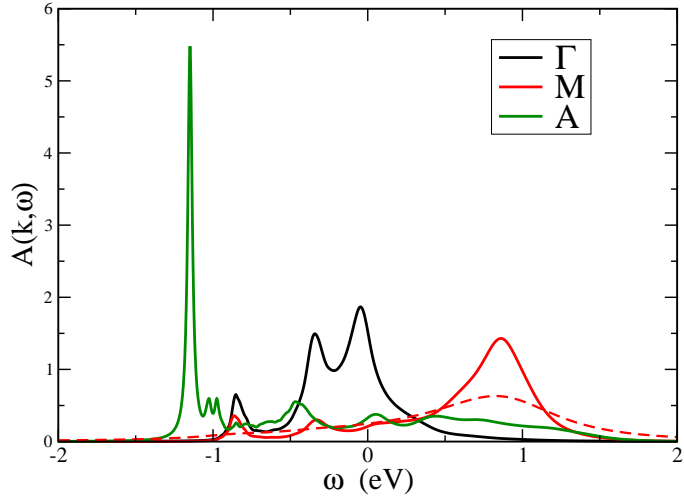}} \hspace{2cm}
\caption{(Color online) Spectral functions corresponding to the $\Gamma$, $M$ and $A$ points, with the hole picture adopted (for comparison with ARPES experiments the electron picture should be adopted). A broadening equivalent to $\sim 20$ meV was applied (see main text). Dashed line: $M$ spectral function with a broadening of $300$ meV. With such a broadening the QP peak is whashed out and only the broad peak near $0.8$ eV persists, which might be misinterpreted as the QP peak.} 
\label{Gamma_M_spectral}
\end{figure*}

An example of the hole spectral function calculated with the SCBA can be seen in Fig. \ref{Gamma_M_spectral} for the $\Gamma$, $M$ and $A$ points. A low broadening, equivalent to $\sim 20$ meV was applied (see main text). For the $\Gamma$ and $M$ points, the quasiparticle weight is low, and most of the spectral weight corresponds to the incoherent part of the spectral function. In cases like these, the quasiparticle energy (Fig. 2 main text) does not coincide with the brighter areas of the intensity map usually plotted in the ARPES experiments (Fig. 3 main text). But if a very high broadening is used, the QP peak is washed out when its weight is low, and hence for these cases the dispersion might be mistakingly shifted to the energy of the incoherent resonances, coinciding with the brighter areas of Fig. 3 in the main text. This is exemplified for the $M$ point in dashed lines in fig \ref{Gamma_M_spectral}. It is clear that the QP energy should be defined with a low broadening.

\end{widetext}


\begin{references}
\bibitem{aclaracion} Note that in the exact diagonalization of the QP dispersion, unless some small anisotropy is put by hand, 
the two magnetic degenerate ground states coexist in the calculation and, hence, the results should reflect this fact \cite{Hamad08}.
\end{references}
\begin{references}
\bibitem{note} To explain some Raman and photoemission experiments at higher
energies, other orbitals should be included (see for example Ref. \onlinecite{raman}), 
but we can neglect them in this work.
\bibitem{note2} This transformation does not correspond to a rotation of the
spins in an angle $\pi $. In particular doing the transformation twice is
equivalent to the identity, while a rotation in $2\pi $ gives minus the
identity.
\end{references}
\end{document}